\documentclass[onecolumn,secnumarabic,amsmath,amssymb,balancelastpage,nofootinbib]{article}

\usepackage[e]{esvect}
\usepackage{color}         
\usepackage{graphics}      
\usepackage{graphicx}      
\usepackage{epsf}          
\usepackage{bm}            

\usepackage{natbib}
\usepackage{amssymb}
\usepackage{amsmath, esint}
\usepackage{mathrsfs}
\usepackage{framed}
\usepackage{bigints} 
\usepackage{enumitem}
\usepackage{pifont}
\usepackage[capbesideposition={left,center},facing=yes,capbesidewidth=8cm,capbesidesep=quad]{floatrow} 
\usepackage{setspace} 

\usepackage[none]{hyphenat} 

\usepackage[colorlinks=true]{hyperref}  

\definecolor{darkred}{rgb}{0.6,0,0}
\definecolor{darkgreen}{rgb}{0,0.5,0}
\definecolor{darkblue}{rgb}{0,0,0.6}
\hypersetup{ colorlinks,
linkcolor=darkblue,
filecolor=darkgreen,
urlcolor=darkgreen,
citecolor=darkred }

\begin{document}

\sloppy 

\bibliographystyle{authordate1}


\title{\vspace*{-35 pt}\huge{Particles, Fields, and the Measurement of\\Electron Spin}}
\author{Charles T. Sebens\\California Institute of Technology}
\date{September 3, 2020\\arXiv v.2\\Forthcoming in \emph{Synthese}}

\maketitle
\vspace*{-20 pt}
\begin{abstract}
This article compares treatments of the Stern-Gerlach experiment across different physical theories, building up to a novel analysis of electron spin measurement in the context of classical Dirac field theory.  Modeling the electron as a classical rigid body or point particle, we can explain why the entire electron is always found at just one location on the detector (uniqueness) but we cannot explain why there are only two locations where the electron is ever found (discreteness).  Using non-relativistic or relativistic quantum mechanics, we can explain both uniqueness and discreteness.  Moving to more fundamental physics, both features can be explained within a quantum theory of the Dirac field.  In a classical theory of the Dirac field, the rotating charge of the electron can split into two pieces that each hit the detector at a different location.  In this classical context, we can explain a feature of electron spin that is often described as distinctively quantum (discreteness) but we cannot explain another feature that could be explained within any of the other theories (uniqueness).
\end{abstract}

\tableofcontents
\newpage

\section{Introduction}\label{introsec}

The measurement of electron spin through the Stern-Gerlach experiment serves as one of the primary touchstone experiments used to introduce and understand quantum physics, vying with the double-slit experiment for placement at the beginning of introductions to the subject.  In philosophical discussions of the foundations of quantum physics, this experiment is ubiquitous.\footnote{Stern-Gerlach spin measurements play a central role in textbook treatments of the philosophy of quantum physics, such as \citet{albertQM, lewisQM, norsen2017, maudlinQM, barrettQM}.\label{textbooks}}  The experiment is ordinarily used to illustrate the ways in which quantum particles differ from classical particles.  Here I will use the experiment to examine the ways in which quantum fields differ from classical fields.

A classical rigid body or point particle electron would be deflected by the inhomogeneous magnetic field of a Stern-Gerlach experiment and could hit anywhere among a continuous range of different locations on the detector.  These classical models of the electron accurately predict that this experiment will have a unique outcome, with the entire electron found at a single location on the detector.  However, these models fail to predict that the outcomes are discrete:\ electrons always hit the detector in one of just two possible locations.\footnote{Stern and Gerlach saw their experiment as demonstrating ``directional quantization,'' though this was not initially applied to electron spin as that idea had not yet entered quantum theory \citep{sterngerlach1922, weinert1995, sauer2016, schmidt2016}.  Now, it is common to present the experiment as showing that electron spin is quantized.  I will use the term ``discrete'' instead of ``quantized'' because ``quantized'' suggests ``quantum'' and, as we will see, the outcomes of Stern-Gerlach experiments can be discrete in theories that are not quantum.}  In non-relativistic quantum mechanics, we can explain this discreteness by analyzing the evolution of the wave function under the Pauli equation.  We can also explain the uniqueness of outcomes, though the way that uniqueness is explained will depend on one's preferred interpretation of quantum mechanics.

Although non-relativistic quantum mechanics is sufficient for explaining the observed results of Stern-Gerlach experiments, it is not our best theory of the electron.  That honor belongs to quantum field theory.  There is debate as to how this theory should be understood, but one approach is to view it as a theory of fields in quantum superpositions of different classical states (just as non-relativistic quantum mechanics is a theory of point particles in quantum superpositions of different classical states).  For the electron, the relevant field is the Dirac field.  To better understand electron spin and the Stern-Gerlach experiment in quantum field theory, we can start by analyzing the experiment in the context of classical Dirac field theory.  To my knowledge, this has not been done before.

Modeling the electron classically using Dirac field theory, we will see that its energy and charge are initially spread out and rotating.  If the electron is rotating about the right axis, then when it passes through the Stern-Gerlach magnetic field it experiences a force and is deflected (behaving just as it would if we modeled it as a rigid body or point particle).  If the electron is rotating about a different axis, then when it passes through the magnetic field it splits into two pieces rotating about the same axis in opposite directions.  The two pieces are each deflected and hit the detector at different locations.  The outcomes of the experiment are discrete but not always unique.  When we quantize classical Dirac field theory and move to quantum field theory, the new quantum feature of these experiments is the uniqueness of outcomes (not the discreteness, which was present already in the classical description).

Although this article can be read on its own, one of its goals is to strengthen the account of electron spin in classical and quantum field theory developed in \citet{howelectronsspin, positrons} (which shares similarities with the accounts in \citealp{ohanian, chuu2010}).  In those articles, I focused on understanding how the electron's angular momentum and magnetic moment can be attributed to the actual rotation of the electron's energy and charge.  But, there was no discussion of Stern-Gerlach experiments or the discrete (two-valued) nature of spin as revealed by such experiments.  This article addresses that gap.  Although there remain some issues---such as the problem of self-repulsion and questions about the role of Grassmann numbers in quantum field theory (see sections \ref{cftsection} and \ref{qftsection})---we will arrive at a clear picture of how electrons behave during Stern-Gerlach experiments according to classical field theory and we will see in outline how one might describe their behavior in quantum field theory.

This article is divided into sections based on the physical theory that is being used to model the electron as it passes through the Stern-Gerlach setup.  We will start with non-relativistic rigid body classical mechanics,\footnote{For philosophical discussion and comparison of rigid body classical mechanics (used in section \ref{rigidbodysection}), point particle classical mechanics (used in section \ref{pointparticlesection}), and continuum mechanics (used in section \ref{cftsection}), see \citet{wilson1998, wilson2013}.} treating the electron as a small rotating sphere of uniform charge density.  Then, we will replace this sphere with a non-relativistic\footnote{We will not examine relativistic classical theories of a point electron in this article (see \citealp{wen2016}; \citealp{barandes2019long, barandes2019short} and references therein).} classical point particle that has intrinsic magnetic moment and angular momentum.  After that, we will switch to a non-relativistic quantum theory of the electron interacting with a classical electromagnetic field via the Pauli equation.  We will then continue on to a relativistic quantum theory where this interaction is modeled by the Dirac equation.  Once this standard story has been retold, highlighting various features and faults, in section \ref{cftsection} we will return to the beginning and start afresh with a different classical theory:\ classical Dirac field theory.  Using this theory, we can describe the flow of charge within the electron and the forces acting on the electron.  We will see that the classical field model of the electron has a number of advantages over the earlier rigid body and point particle models.  The following section discusses how one would model the electron in the context of quantum field theory, but---because of the complexity of this context---we will not go into the same level of mathematical detail as in the preceding sections.  Using each of these models of the electron, we will repeatedly analyze the same two example cases:\ $z$-spin up and $x$-spin up electrons in a Stern-Gerlach experiment oriented to measure the $z$ component of the electron's spin.  The use of the word ``measure'' in the previous sentence fits with ordinary usage but is not entirely accurate.  As will be explained in the appropriate sections, the Stern-Gerlach experiment only truly acts as a measurement of the $z$ component of the electron's spin in the first two classical contexts:\ classical rigid body mechanics and classical point particle mechanics.

\section{Classical Rigid Body Mechanics}\label{rigidbodysection}

\begin{figure}[p!]
\center{\includegraphics[width=10 cm]{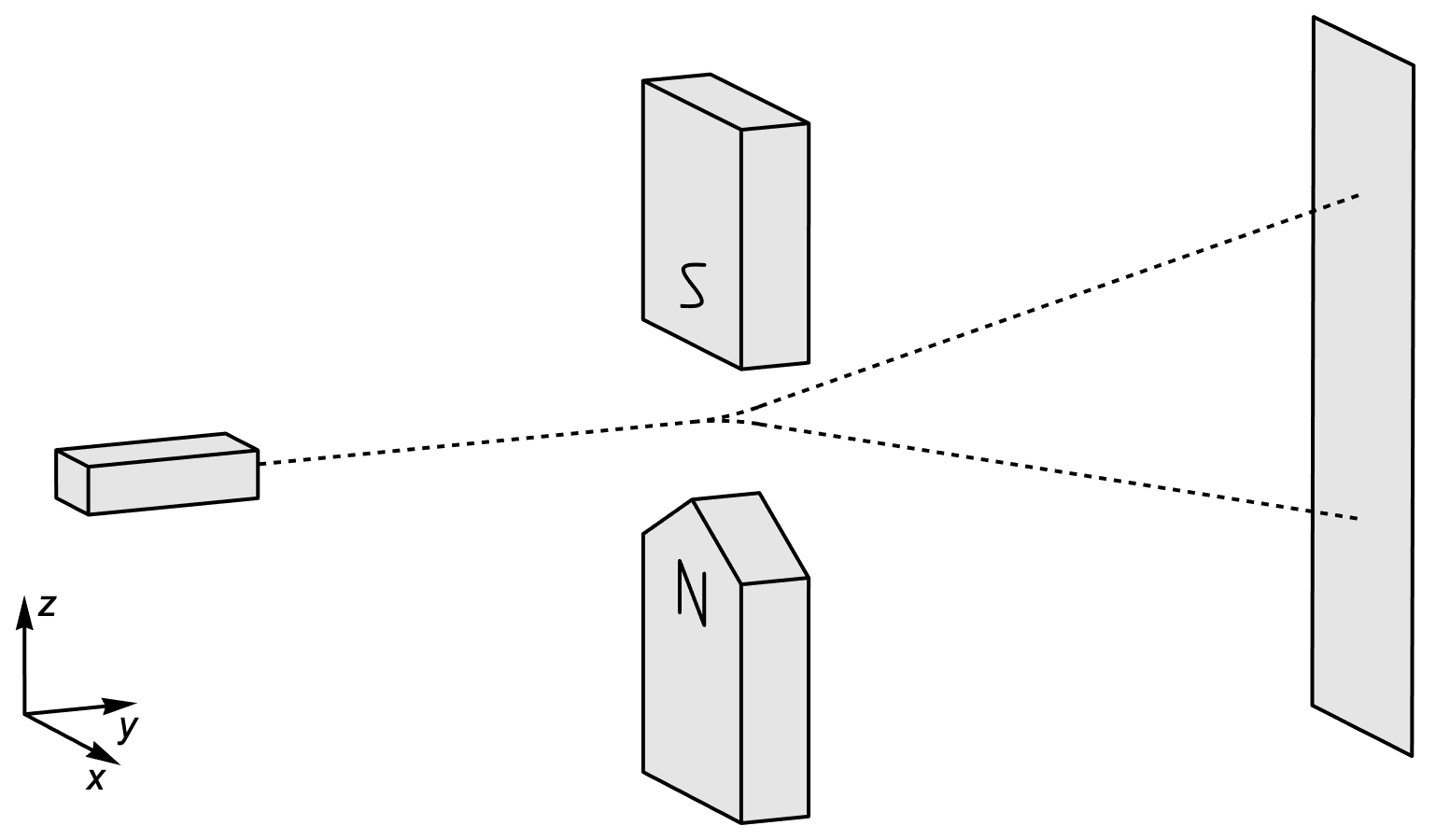}}
\caption{In the Stern-Gerlach experiment, electrons are shot from an emitter (on the left) through an inhomogeneous magnetic field produced by two magnets (in the center), after which their final locations are recorded when they hit a detector screen (on the right).  The dotted lines show the paths of $z$-spin up and $z$-spin down electrons.}
  \label{SGsetup}
\end{figure}

Consider a Stern-Gerlach measurement of an electron's spin along the $z$ axis.  Suppose that the electron begins moving in the $y$ direction, is deflected as it passes through an inhomogeneous magnetic field, and then hits a detector (depicted in figure \ref{SGsetup}).  One might try to calculate its path using the Lorentz force law for point particles:\footnote{Here and throughout I adopt Gaussian cgs units.}
\begin{equation}
\vec{F}= q  \vec{E} + \frac{q}{c} \vec{v} \times \vec{B}
\ ,
\label{pforcelaw}
\end{equation}
where $q$ is the particle's charge and $\vec{v}$ is its velocity.  As there is no electric field acting on the electron, the first term is zero.  The second term captures the deflection resulting from the fact that the electron is a charged particle moving through a magnetic field---a force that points in the $x$ direction when the electron first hits the inhomogeneous field.  This force is independent of the orientation of the electron's magnetic moment.  It is not the deflection that the Stern-Gerlach apparatus is designed to measure and it is not shown in figure \ref{SGsetup}.  In practice, this complication is generally removed by not sending individual negatively charged electrons through this gauntlet, but instead uncharged atoms (originally silver) where most of the electrons have their magnetic moments paired (so that they cancel) except for one electron that has an unpaired magnetic moment.  For our theoretical purposes, we can just acknowledge the existence of this sideways deflection and put it aside to focus on other contributions to the total deflection.

\begin{figure}[p!]
\center{\includegraphics[width=7 cm]{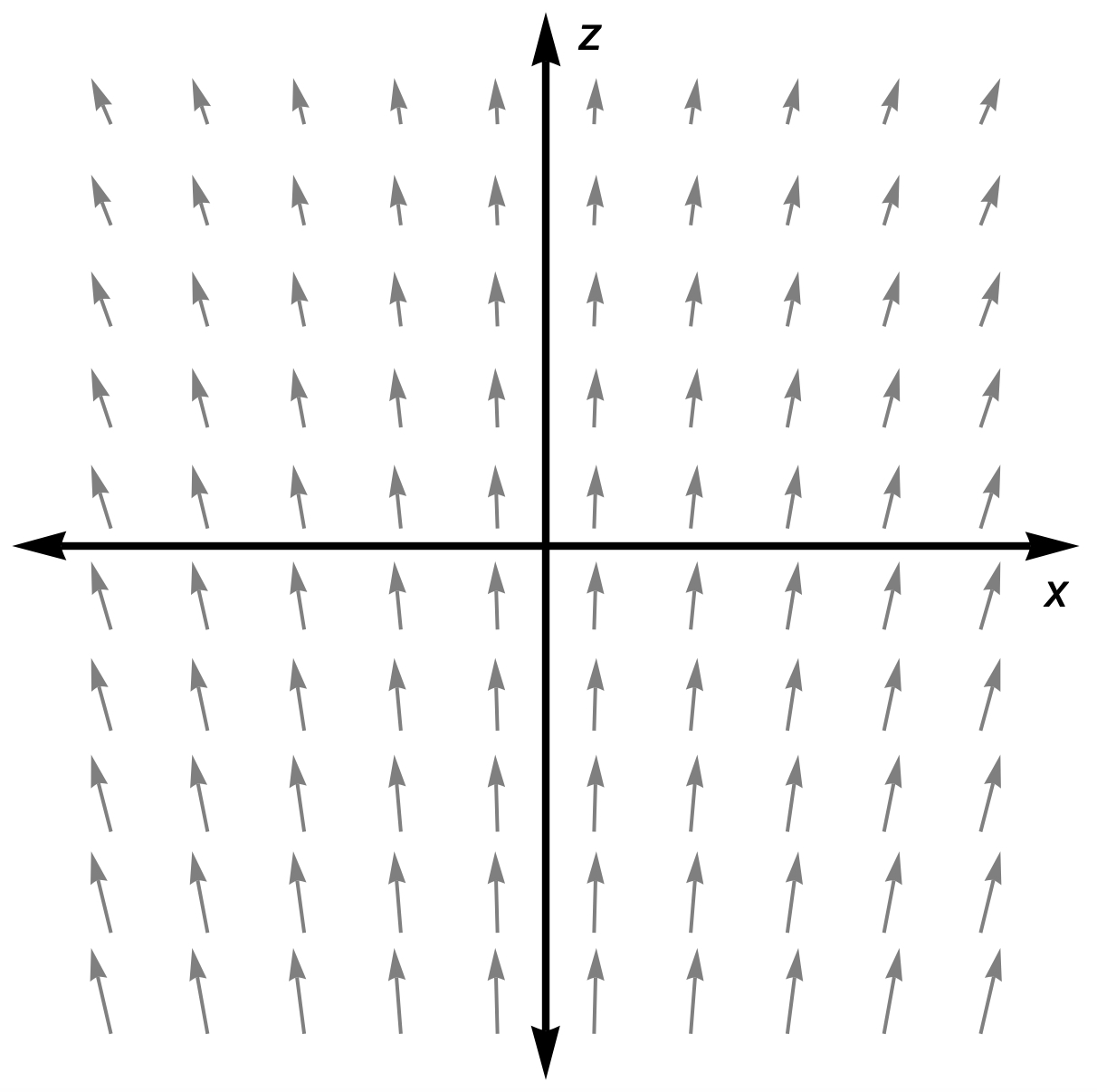}}
\caption{The magnetic field of the Stern-Gerlach experiment \eqref{SGfield} points primarily upwards, becoming weaker as we move up in the $z$ direction and tilting outwards as we move away from the $z$ axis.}
  \label{SGfieldfig}
\end{figure}

The force that is most important to our analysis of the Stern-Gerlach experiment is the force that deflects the electron upwards or downwards, depending on the orientation of its magnetic moment.  This force does not appear in the above point particle Lorentz force law \eqref{pforcelaw}.  We can calculate this force by beginning instead with the Lorentz force law for distributions of charge:
\begin{equation}
\vec{f}= \rho^q  \vec{E} + \frac{1}{c} \vec{J} \times \vec{B}
\ ,
\label{cforcelaw}
\end{equation}
where $\vec{f}$ is the force density, $\rho^q$ is the charge density, and $\vec{J}$ is the current density.  Let us model the electron (for practical convenience) as a rigid sphere with radius $R$, total mass $m$, and total charge $-e$ (with the mass and charge distributed evenly throughout the electron's volume).  Further, let us assume that the electron is $z$-spin up and that the electron's charge is rotating uniformly, assigning it a current density (within the sphere) of
\begin{equation}
\vec{J}=\frac{15 \mu c}{4 \pi R^5} ( \vec{x}\times \hat{z} )
\ ,
\label{zupcurrentsphere}
\end{equation}
so that the magnetic moment\footnote{For the definition of magnetic moment in terms of current density, see \citet[sec.\ 5.6]{jackson}.} of the electron is the Bohr magneton, $\mu=\frac{e \hbar}{2 m c}$,
\begin{equation}
\vec{m}=\frac{1}{2c}\int dV \left(\vec{x} \times \vec{J}\,\right)=-\mu \hat{z}
\ .
\label{zupmagneticmoment}
\end{equation}
In these equations, $\hat{z}$ is the unit vector pointing in the $z$ direction.  We can similarly assign the electron a momentum density (within the sphere) of
\begin{equation}
\vec{G}=-\frac{15 \hbar}{16 \pi R^5} ( \vec{x}\times \hat{z} )
\ ,
\label{zupmomentumsphere}
\end{equation}
so that its total angular momentum is
\begin{equation}
\vec{L}=\int dV \left(\vec{x} \times \vec{G}\right)=\frac{\hbar}{2}\hat{z}
\ .
\label{zupangularmomentum}
\end{equation}
It is because this angular momentum points upwards that the electron is called ``$z$-spin up'' (even though the magnetic moment points downwards).

In the context of classical rigid body mechanics, there is no explanation as to why the spherical electron is rotating as described above instead of a bit faster or slower.  For now, we can leave this unexplained and proceed by positing fixed values for the electron's angular momentum and magnetic moment.\footnote{Keeping the electron's angular momentum and magnetic moment fixed, the electron will have to rotate faster the smaller it is.  For a sufficiently small radius, its edges would have to move faster than the speed of light (\citealp[pg.\ 35]{tomonaga1997}; \citealp[problem 4.25]{griffithsQM}; \citealp[pg.\ 127]{rohrlich}; \citealp{howelectronsspin}).  Let us assume that the electron is large enough that it does not have to rotate so rapidly.}  These values cannot be obtained by a spherical body with a uniform distribution of mass and charge rotating together at some particular angular velocity.  That kind of body would have a different gyromagnetic ratio (the ratio of magnetic moment to angular momentum).\footnote{See \citet[pg.\ 47]{uhlenbeck}; \citet[pg.\ 39]{pais1989}; \citet[pg.\ 187]{jackson}; \citet[problem 5.58]{griffiths}.}  However, we can allow the angular momentum and magnetic moment in \eqref{zupmagneticmoment} and \eqref{zupangularmomentum} if we do not require that the mass and charge always rotate together at the same rate.\footnote{See \citet{howelectronsspin}.}

In the above equations, the electron as a whole is not moving.  One could give it a velocity in the $y$ direction so that it passes through the inhomogeneous magnetic field of the Stern-Gerlach apparatus and then hits the detector.  We will simplify the calculations by choosing a frame in which the electron rotates in place (as described above) and the magnets move towards it, modeling the effect of the magnets by briefly turning on the inhomogeneous magnetic field and seeing how the electron reacts.  In this frame, the sideways deflection that was previously attributed to the electron being a charged particle moving through a magnetic field (and set aside) would instead be attributed to the electron at rest responding to the electric field induced by the moving magnets.  Because we are not particularly interested in this deflection, we will not analyze that electric field (in this section or in those that follow).

In the Stern-Gerlach experiment, the precise form of the inhomogeneous magnetic field will depend on the shapes of the magnets and their arrangement.  For simplicity, we will take the field between the magnets to be the sum of a strong background field $B_0\hat{z}$ and an inhomogeneity $\eta x \hat{x}- \eta z\hat{z}$ (where $\eta$ is a positive constant characterizing the strength of the inhomogeneity),\footnote{This expression for the magnetic field follows \citet[pg.\ 181]{griffithsQM}.}
\begin{equation}
\vec{B}_{SG}=\eta x \hat{x}+(B_0 - \eta z)\hat{z}
\ .
\label{SGfield}
\end{equation}
This field is shown in figure \ref{SGfieldfig}.  As it will be relevant later, note that we can describe this magnetic field via the vector potential
\begin{equation}
\vec{A}_{SG}=(B_0 x - \eta x z)\hat{y}
\ ,
\label{SGvector}
\end{equation}
related to the magnetic field via $\vec{B}=\vec{\nabla}\times\vec{A}$.

\begin{figure}[p!]
\center{\includegraphics[width=12cm]{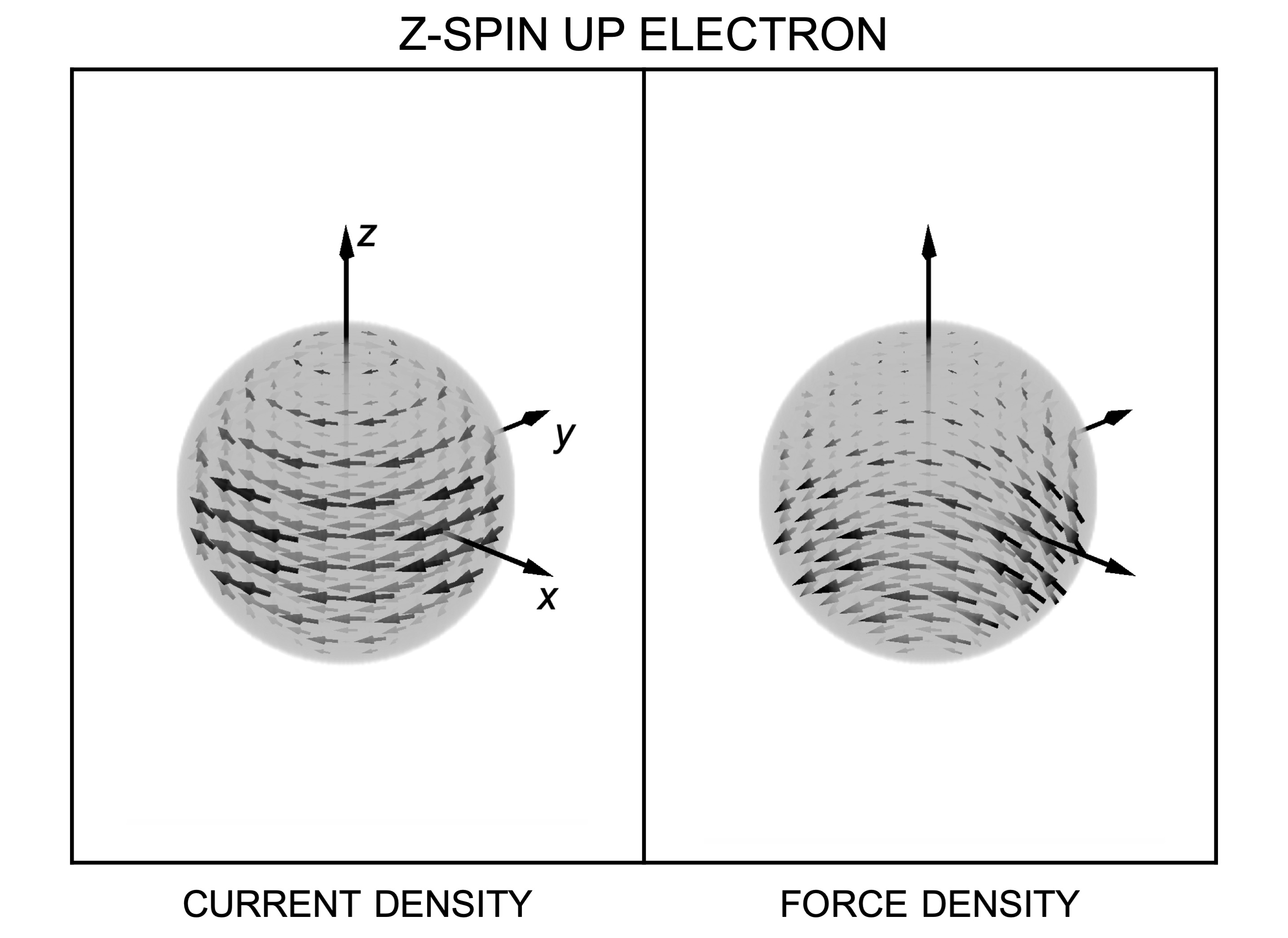}}\\
\center{\includegraphics[width=12cm]{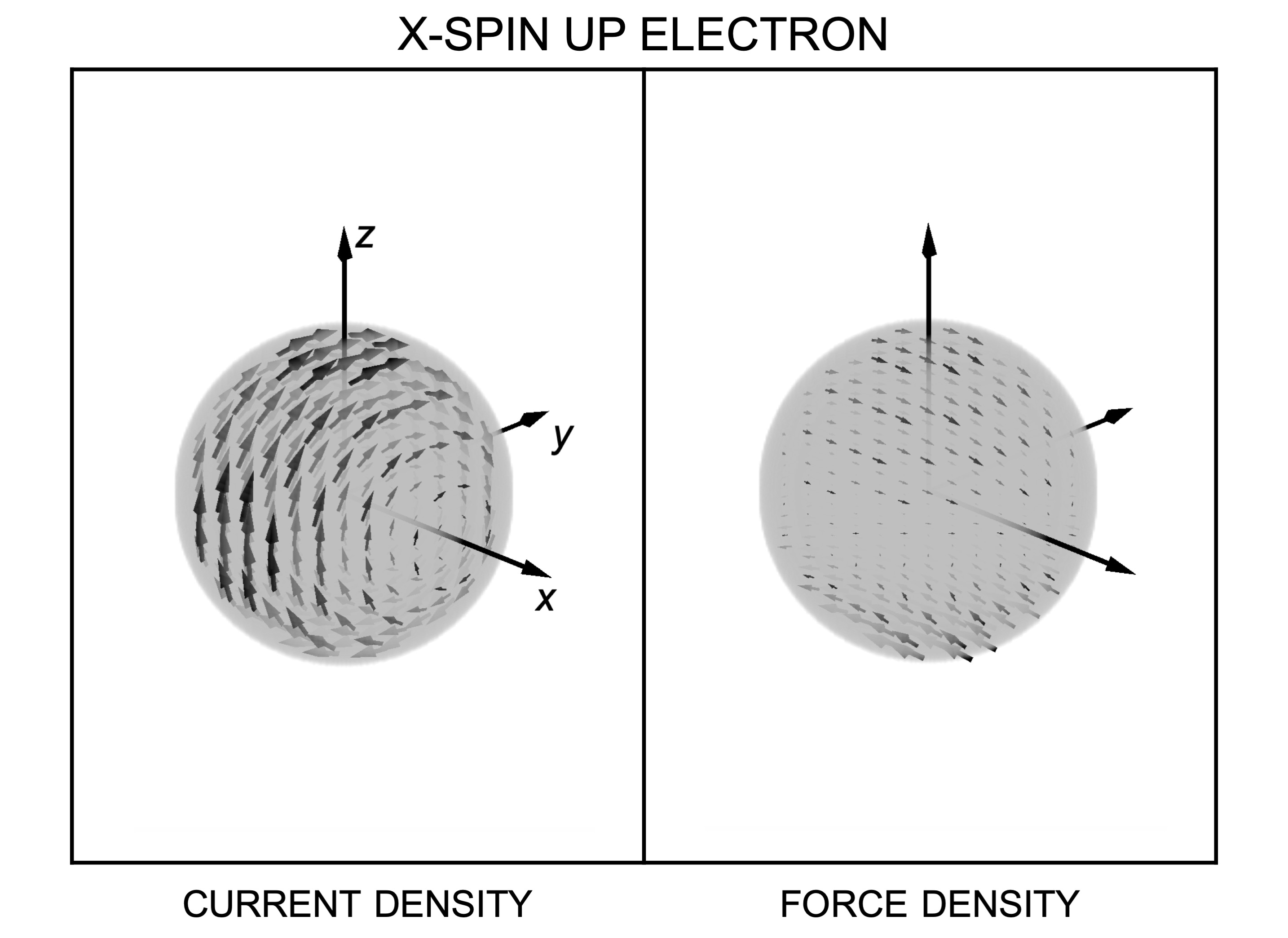}}
\caption{These plots show the initial current and force densities of $z$-spin up and $x$-spin up classical spherical electrons entering the magnetic field of the Stern-Gerlach experiment---\eqref{zupcurrentsphere}, \eqref{zupsphereforcedensity}, \eqref{xupcurrentsphere}, and \eqref{xupsphereforcedensity}.  Each plot shows the charge density as a uniform gray cloud and the aforementioned densities as arrows of varying sizes (plotted at points on a grid).  The arrows that you can see most clearly are near the surface as many are either partially or fully obscured by the cloud (reducing clutter).  In plotting the force densities, the variation in the magnetic field across the diameter of the electron has been exaggerated so that it is easy to see details like the force pulling the $z$-spin up electron upwards (visible on the right side of the first force density).}
  \label{SG1}
\end{figure}

To see why the $z$-spin up electron is deflected upwards as it passes through this magnetic field, we can calculate the density of force on the electron using \eqref{cforcelaw}, \eqref{zupcurrentsphere}, and \eqref{SGfield},
\begin{align}
\vec{f}&=\frac{1}{c}\vec{J}\times\vec{B}_{SG}
\nonumber
\\
&=\frac{15 \mu}{4 \pi R^5} (\vec{x}\times\hat{z})\times(\eta x \hat{x} +(B_0 - \eta z)\hat{z})
\nonumber
\\
&=\frac{15 \mu}{4 \pi R^5} \left( (-B_0 x + \eta x z) \hat{x} + (B_0 y - \eta y z)\hat{y} + \eta x^2 \hat{z} \right)
\ .
\label{zupsphereforcedensity}
\end{align}
This force density is shown in figure \ref{SG1}.  The $x$ and $y$ components of the force density give no net force and no net torque when integrated over the sphere, so they can be neglected for our purposes.  The  $z$ component of the force density (which comes from the $x$ component of $\vec{B}_{SG}$) is responsible for sending the electron upwards.  This force density is zero when $x=0$ and strongest at the electron's sides where $x^2$ is largest.  Thinking of the electron as a head facing in the $y$ direction (towards the detector), it is pulled up by its ears.  We can find the total force exerted on the electron by integrating the $z$ component of the force density over the volume of the sphere,
\begin{equation}
\vec{F}= \int dV \left( \frac{15 \mu \eta}{4 \pi R^5} x^2 \hat{z}\right)=\mu \eta \hat{z}
\ .
\label{zforcesphere}
\end{equation}
If the electron is in the Stern-Gerlach magnetic field for a time $\Delta t$, then it will acquire a momentum of
\begin{equation}
\vec{p}=\mu \eta \Delta t \hat{z}
\label{zmomentumsphere}
\end{equation}
and move upwards with a velocity of
\begin{equation}
\vec{v}=\frac{\mu \eta \Delta t}{m} \hat{z}
\ .
\label{zvelocitysphere}
\end{equation}

We can calculate the energy associated with the electron's presence in the Stern-Gerlach magnetic field by analyzing the energy density of the electromagnetic field,
\begin{equation}
\frac{E^2}{8 \pi} + \frac{B^2}{8 \pi}
\ ,
\label{EMenergydensity}
\end{equation}
and seeing how the total energy in the electromagnetic field changes when the electron enters the magnetic field.  Considering spinning sphere states like \eqref{zupcurrentsphere} and allowing the location of the electron's center and the orientation of its magnetic moment to vary, one can derive the following general expression for this change in field energy:
\begin{equation}
- \vec{m} \cdot \vec{B}_{SG}
\ .
\label{energyder2}
\end{equation}
Because the additional energy in the electromagnetic field depends on the electron's location in the field, one might think of the energy in \eqref{energyder2} as a potential energy of the electron (as opposed to an actual energy of the field).  Attributing such energy to the electromagnetic field has the advantage that the expression for the field energy in \eqref{EMenergydensity} ensures conservation of energy in electromagnetic interactions, whereas attributing potential energy to matter does not.\footnote{See \citet[sec.\ 15.6]{feynman2}; \citet[ch.\ 5]{lange}}  That being said, thinking of \eqref{energyder2} as a potential energy allows us to calculate the force on an electron from this potential energy in the standard way,
\begin{align}
\vec{F}&= - \vec{\nabla} U
\nonumber
\\
&= \vec{\nabla} \left(\vec{m}\cdot\vec{B}_{SG} \right)
\ ,
\label{forceder}
\end{align}
where $U$ is the position-dependent potential energy of the electron.  For a $z$-spin up electron, the force is $\mu \eta \hat{z}$.  This result is in agreement with the force calculated earlier \eqref{zforcesphere}.  In \eqref{forceder} the force on a $z$-spin up spherical electron appears to originate from the $- \eta z\hat{z}$ term in the magnetic field \eqref{SGfield}.  In \eqref{zupsphereforcedensity} and \eqref{zforcesphere} we saw that the Lorentz force on such an electron actually originates from the $\eta x \hat{x}$ term in the magnetic field.

By repeating the earlier application of the Lorentz force law \eqref{cforcelaw}, we can calculate the density of force on an electron with any spin orientation.  For our purposes, it will be illustrative to consider an electron that is $x$-spin up.  For such an electron, the current density would be
\begin{equation}
\vec{J}=\frac{15 \mu c}{4 \pi R^5} ( \vec{x}\times \hat{x} )
\ ,
\label{xupcurrentsphere}
\end{equation}
as in \eqref{zupcurrentsphere} but with the current rotating about the $x$ axis instead of the $z$ axis.  When such an electron enters the magnetic field of the Stern-Gerlach apparatus \eqref{SGfield}, the force density it experiences can be calculated using the Lorentz force law \eqref{cforcelaw}, as in \eqref{zupsphereforcedensity},
\begin{align}
\vec{f}&=\frac{15 \mu}{4 \pi R^5} (\vec{x}\times\hat{x})\times(\eta x \hat{x} +(B_0 - \eta z)\hat{z})
\nonumber
\\
&=\frac{15 \mu}{4 \pi R^5} \left( (B_0 z - \eta z^2) \hat{x} - \eta x y \hat{y} - \eta x z \hat{z} \right)
\ .
\label{xupsphereforcedensity}
\end{align}
It is clear that along the $z$ axis there is no force and thus that this $x$-spin up electron will be deflected neither upwards nor downwards.  However, because of the  $- \eta z^2 \hat{x}$ term, it appears that the electron will be deflected sideways, in the minus $x$ direction---and that an $x$-spin down electron would feel an opposite force and be deflected in the positive $x$ direction.  But, the Stern-Gerlach setup was just supposed to sort electrons base on the $z$ components of their magnetic moments (not the $x$ components).  This is where the strong homogenous component of the magnetic field becomes very important.  It exerts a torque on the electron that causes its magnetic moment to precess about the $z$ axis (Larmor precession\footnote{See \citet[sec.\ 34.3]{feynman2}; \citet[sec.\ 45]{landaulifshitzfields}; \citet{alstrom1982}; \citet[sec.\ 4.4.2]{griffithsQM}.}).  The existence of such a torque is apparent in \eqref{xupsphereforcedensity} and the illustration of the force density in figure \ref{SG1}.  The torque on the electron is initially,
\begin{align}
\vec{\tau}&=\int d^3 x \ \vec{x} \times \vec{f}
\nonumber
\\
&=\mu B_0 \hat{y}
\ .
\label{xupspheretorque}
\end{align}
The $B_0 z \hat{x}$ term will impart some angular momentum about the $y$ axis, shifting the electron's total angular momentum vector from pointing squarely in the $x$ direction to point partially in the $y$ direction.  In general (allowing the electron's location and spin orientation to vary), the torque on the electron is $\vec{\tau}= \vec{m}\times\vec{B}_{SG}$.  If we suppose that the electron's angular momentum and magnetic moment always remain oppositely directed, then from the form of that equation it is clear that both will precess about the $z$ axis with an angular frequency of $\frac{2\mu}{\hbar} B_0$ (assuming $\eta$ is small and the electron remains near the origin).

Because the orientation of an electron's magnetic moment is rapidly changing as the electron undergoes this Larmor precession, the net force exerted over a period of time by the Stern-Gerlach magnetic field on an electron depends only on the $z$ component of its magnetic moment.  For an electron that starts $x$-spin up, the force is essentially zero.  Such an electron will not be deflected as it passes through the Stern-Gerlach apparatus.  Of course, when the experiment is actually conducted, electrons are always either deflected (maximally) upwards or (maximally) downwards.  However, if we model the electron using classical rigid body mechanics, it should be able to hit anywhere in between (depending on the orientation of its spin).  This shortcoming is ordinarily resolved by moving to a quantum treatment of spin built from a classical theory where the electron is modeled as a point particle.  Let us examine that classical theory next (and then move on to quantum theories).

\section{Classical Point Particle Mechanics}\label{pointparticlesection}

At the beginning of the previous section, we saw that if we model the electron as a point particle obeying the Lorentz force law, we cannot account for the crucial deflection of the electron in the Stern-Gerlach magnetic field.  However, if we model the electron as a rigid sphere then we can use the Lorentz force law on its own to calculate these upwards and downwards forces.  We can also use that force law to calculate torques that lead to Larmor precession (and keep electrons from being deflected sideways).  It is possible to incorporate these effects into a theory of point particles, but it comes at a cost to the simplicity, elegance, and unity of our physics:\ (i) we must endow these point particles with intrinsic magnetic moments and (oppositely oriented) intrinsic angular momenta,\footnote{The spherical electron of section \ref{rigidbodysection} had a magnetic moment as well, but it was not intrinsic.  It was the result of flowing charge.  Similarly, the spherical electron had angular momentum that was the result of flowing mass.} (ii) we must modify the Lorentz force law to include a term that depends on the magnetic moments (which will be responsible for the key deflection in the Stern-Gerlach experiment), (iii) we must introduce a new equation specifying the torque on these point size particles (which will be necessary to account for Larmor precession).\footnote{Although we can derive the additional term in the force law that must be postulated and the equation for the torque that must be added by analyzing the theory of electromagnetism with rigid bodies, these features must be included as part of the fundamental dynamical laws in this theory of electromagnetism with point particles.}  By contrast, the model in the previous section only had flowing mass and charge (no intrinsic angular momenta or magnetic moments) evolving in accord with the Lorentz force law.  Looking at these costs, I think a model of the electron like the one in the previous section is attractive and later, in section \ref{cftsection}, I will put forward a classical model that is in many ways similar.  However, for the time being, let us march on with the standard story and develop a point particle model of electron spin.

In this section we will model the electron as a point particle with mass $m$, charge $-e$, intrinsic magnetic moment $\vec{m}$ of magnitude $\mu$, and intrinsic angular momentum $\vec{L}$ of magnitude $\frac{\hbar}{2}$ (with $\vec{m}$ and $\vec{L}$ always pointing in opposite directions).  The Lorentz force law can be extended from \eqref{pforcelaw} to capture all of the forces acting on a particle that carries both charge and intrinsic magnetic moment,\footnote{\citet[pg.\ 378]{griffiths} considers this modification to the Lorentz force law to incorporate intrinsic magnetic moment and remarks: ``I don't know whether a consistent theory can be constructed in this way, but in any event it is \emph{not} classical electrodynamics, which is predicated on Amp\`{e}re's assumption that all magnetic phenomena are due to electric charges in motion, and point magnetic dipoles must be interpreted as the limits of tiny current loops.''  \citet{barandes2019long, barandes2019short} analyzes such modifications to the force law, considering intrinsic magnetic dipole moments, intrinsic electric dipole moments, and also other multipole moments.\label{modifiedforcefootnote}}
\begin{equation}
\vec{F}= q  \vec{E} + \frac{q}{c} \vec{v} \times \vec{B} + \frac{1}{c}\vec{\nabla} \left(\vec{m}\cdot\vec{B} \right)
\ .
\label{pforcelaw2}
\end{equation}
As was discussed regarding \eqref{forceder}, the force described by the third term in \eqref{pforcelaw2} might be interpreted as the negative gradient of a potential energy\footnote{For a point electron, this potential energy cannot be derived from the total energy in the electromagnetic field (as in section \ref{rigidbodysection}).  Instead, this potential energy can be calculated by considering the work that must be done to rotate a point magnetic dipole in an external magnetic field \citep[problem 6.21]{griffiths} or the work required to bring the dipole to a given location from infinity \citep[pg.\ 227]{goodnelson1971}.  Or, it can be calculated by asking what potential would generate the additional term in the modified Lorentz force law \eqref{pforcelaw2} when one takes its negative gradient \citep[sec.\ 5.7]{jackson}.} for the electron in the magnetic field of the Stern-Gerlach magnets,
\begin{equation}
U=- \vec{m} \cdot \vec{B}_{SG}
\ .
\label{penergy}
\end{equation}
Although this point electron's magnetic moment and angular momentum never change in strength, they can change in direction.  This change is governed by another law\footnote{For a general derivation from the theory of electromagnetism with rigid bodies of the law for torque that should be included in a theory with point bodies that possess intrinsic magnetic moments and angular momenta, see \citet[sec.\ 5.7]{jackson}.} giving the torque exerted by a magnetic field,
\begin{equation}
\vec{\tau}= \vec{m}\times\vec{B}_{SG}
\ .
\label{torquelaw}
\end{equation}
From the form of this equation, it is clear that the torque will always be orthogonal to the electron's intrinsic magnetic moment and angular momentum (and thus that the magnitude of the electron's intrinsic angular momentum will neither increase nor decrease).

We can analyze the Stern-Gerlach experiment as before by imagining an electron that briefly experiences the magnetic field in \eqref{SGfield}.  From the last term in \eqref{pforcelaw2}, we see that a $z$-spin up electron (with $\vec{m}=-\mu\hat{z}$) will experience an upward force of
\begin{equation}
\vec{F}=\mu \eta \hat{z}
\label{zforcepoint}
\end{equation}
when it is in the magnetic field, as in \eqref{zforcesphere}.  In section \ref{rigidbodysection}, the force on the electron originated from the $\eta x \hat{x}$ term in the magnetic field \eqref{SGfield} and was concentrated on the outer edges of the electron (pulling it up by its ears).  In this section, the force on the electron originates instead from the $- \eta z\hat{z}$ term in the magnetic field and is concentrated entirely at the single point where the electron is located.  It would be possible to create a figure like figure \ref{SG1} illustrating the state of the electron before passing through the Stern-Gerlach magnetic field and the force on the electron in that magnetic field, but there would not be much to see.  The point electron's initial magnetic moment points downwards and the force on the electron points upwards.

From the expression for torque in \eqref{torquelaw}, we can see that an $x$-spin up electron will undergo Larmor precession in the magnetic field and that the total force from the last term in \eqref{pforcelaw2} will be essentially zero.  Such an electron would not be deflected by the Stern-Gerlach magnets.

As in the last section, we have a model of the electron where the Stern-Gerlach experiment will have a single unique outcome but where that outcome could lie anywhere between maximal upward deflection and maximal downward deflection (depending on the electron's initial spin orientation).  It is now time to consider a theoretical context that is actually capable of accurately predicting the results of Stern-Gerlach experiments:\ non-relativistic quantum mechanics.

\section{Non-Relativistic Quantum Mechanics}\label{nrqmsection}

In non-relativistic quantum mechanics, the point electron from the previous section enters a superposition of different spin orientations and locations.  The physical state of the electron is specified (at least in part) by a two-component wave function $\chi(\vec{x})$, where the first component assigns a complex amplitude to the point electron being at $\vec{x}$ with $z$-spin up and the second component assigns a complex amplitude to the point electron being at $\vec{x}$ with $z$-spin down.\footnote{When the electron's location is not relevant, its spin state is sometimes just specified by two complex numbers.  We will not analyze the electron using that simplified representation in this article.}  This wave function evolves by the Pauli equation,\footnote{See \citet[eq.\ 1.34]{bjorkendrell}; \citet[eq.\ 33.7]{lifshitzRQM}; \citet[eq.\ 1.28]{durr2020}.}
\begin{equation}
i \hbar \frac{\partial \chi}{\partial t} = \Big( \underbrace{\frac{- \hbar^2}{2 m}\nabla^2}_{\widehat{H}_0}+\underbrace{\mu\: \vec{\sigma} \cdot \vec{B}_{SG}}_{\widehat{H}_I} \Big)\chi
\ ,
\label{pauli}
\end{equation}
where $\widehat{H}_0$ is the free Hamiltonian operator and $\widehat{H}_I$ is an interaction Hamiltonian.  In this version of the Pauli equation, we simplify the interaction between the electron's quantum wave function and the external classical electromagnetic field by only including the one term that is necessary to account for Larmor precession and the deflection of $z$-spin up and $z$-spin down electrons in the Stern-Gerlach experiment.  The interaction Hamiltonian in \eqref{pauli} corresponds to the energy in \eqref{penergy}, where the quantum operator for the electron's magnetic moment is $-\mu \vec{\sigma}$.  The operator for the electron's angular momentum is $\frac{\hbar}{2}\vec{\sigma}$.  Both of these operators are expressed in terms of the Pauli spin matrices,
\begin{equation}
\sigma_x=\left(\begin{matrix} 0 & 1 \\  1 & 0 \end{matrix}\right)
\quad\quad
\sigma_y=\left(\begin{matrix} 0 & -i \\  i & 0 \end{matrix}\right)
\quad\quad
\sigma_z=\left(\begin{matrix} 1 & 0 \\ 0 & -1 \end{matrix}\right)
\ .
\label{matrixdefs}
\end{equation}

As in the last two sections, we'll begin by considering a $z$-spin up electron.  Let us suppose that, immediately before the electron enters the Stern-Gerlach magnetic field, the electron's wave function is a gaussian wave packet centered at the origin,
\begin{equation}
\chi(\vec{x}) = \left(\frac{1}{\pi d^2}\right)^{3/4}  \exp\left[\frac{-|\vec{x}|^2}{2 d^2}\right] \left(\begin{matrix} 1\\0 \end{matrix}\right)
\ ,
\label{zupNRQM}
\end{equation}
where the constant $d$ determines how widely this wave packet is spread.  This wave function has two components, but the second is everywhere zero because the electron is $z$-spin up.  The factor preceding the exponential in \eqref{zupNRQM} can be calculated by requiring the wave function to be normalized.  As in the other sections, we are working in the electron's rest frame (where the expectation value of the electron's momentum is zero).

Using the Pauli equation, we can calculate the time evolution of this state while the electron is in the Stern-Gerlach magnetic field.  As an approximation, let us evolve the wave function over the short time interval $\Delta t$ using only the interaction Hamiltonian and then calculate the subsequent evolution using the free Hamiltonian (simplifying our calculations by ignoring the motion and spread of the wave packet that would result from the free Hamiltonian during the brief period when the electron is in the inhomogeneous magnetic field of the Stern-Gerlach magnets).\footnote{This approximation is used in \citet[example 4.4]{griffithsQM}; \citet[sec.\ 9.1]{ballentine}; \citet[sec.\ 1.7.1]{durr2020}.}  Considering only the interaction Hamiltonian and inserting the Stern-Gerlach field \eqref{SGfield}, the Pauli equation becomes
\begin{equation}
i \hbar \frac{\partial \chi}{\partial t} = \left\{ \mu(B_0-\eta z) \left(\begin{matrix} 1 & 0 \\ 0 & -1 \end{matrix}\right) + \mu \eta x \left(\begin{matrix} 0 & 1 \\  1 & 0 \end{matrix}\right)\right\}\chi
\ .
\label{SGpauli}
\end{equation}
Looking at this evolution for the wave function in \eqref{zupNRQM}, it might initially appear that the $\mu \eta x \sigma_x$ term (coming from the $\eta x \hat{x}$ component of the magnetic field) will lead to the development of a non-negligible $z$-spin down component in $\chi$.  However, the $B_0 \hat{z}$ piece of the magnetic field will make the phase of the upper component of $\chi$ (the $z$-spin up component) oscillate rapidly so that the net effect of the $\mu \eta x \sigma_x$ term in \eqref{SGpauli} can be ignored.\footnote{See \citet{platt1992}.}  Dropping that term, the time evolution of $\chi$ in \eqref{zupNRQM} is given by
\begin{equation}
i \hbar \frac{\partial \chi}{\partial t} = \mu(B_0-\eta z) \chi
\ ,
\end{equation}
and it is straightforward to determine the state after $\Delta t$,
\begin{equation}
\chi(\vec{x}) = \left(\frac{1}{\pi d^2}\right)^{3/4}  \exp\left[\frac{-|\vec{x}|^2}{2 d^2} - \frac{i}{\hbar}\mu B_0 \Delta t + \frac{i}{\hbar}\mu \eta z \Delta t\right] \left(\begin{matrix} 1\\0 \end{matrix}\right)
\ .
\label{zupNRQMafter}
\end{equation}
The $-\frac{i}{\hbar}\mu B_0 \Delta t$ term in the exponential is irrelevant as it corresponds to the rapid phase oscillation mentioned earlier, capturing only where the roulette wheel happened to stop.  This overall phase factor is independent of location and does not affect the future time evolution of the wave packet.  The $\frac{i}{\hbar}\mu \eta z \Delta t$ term, on the other hand, is critical.  This $z$ dependent phase oscillation has given our wave packet a non-zero momentum in the $z$ direction.  As a consequence, the wave packet will move upwards as it evolves under the free Hamiltonian. The expectation value of the momentum operator for the wave function in \eqref{zupNRQMafter} is
\begin{equation}
\int d^3 x \ \chi^{\dagger} \left( -i \hbar \vec{\nabla} \right) \chi = \mu \eta \Delta t \hat{z}
\ .
\label{momentumexpectation}
\end{equation}
The new momentum of the electron matches the classical prediction from the forces on a rigid sphere \eqref{zforcesphere} or a point electron \eqref{zforcepoint}, given in \eqref{zmomentumsphere}.

\begin{figure}[p!]
\center{\includegraphics[width=12cm]{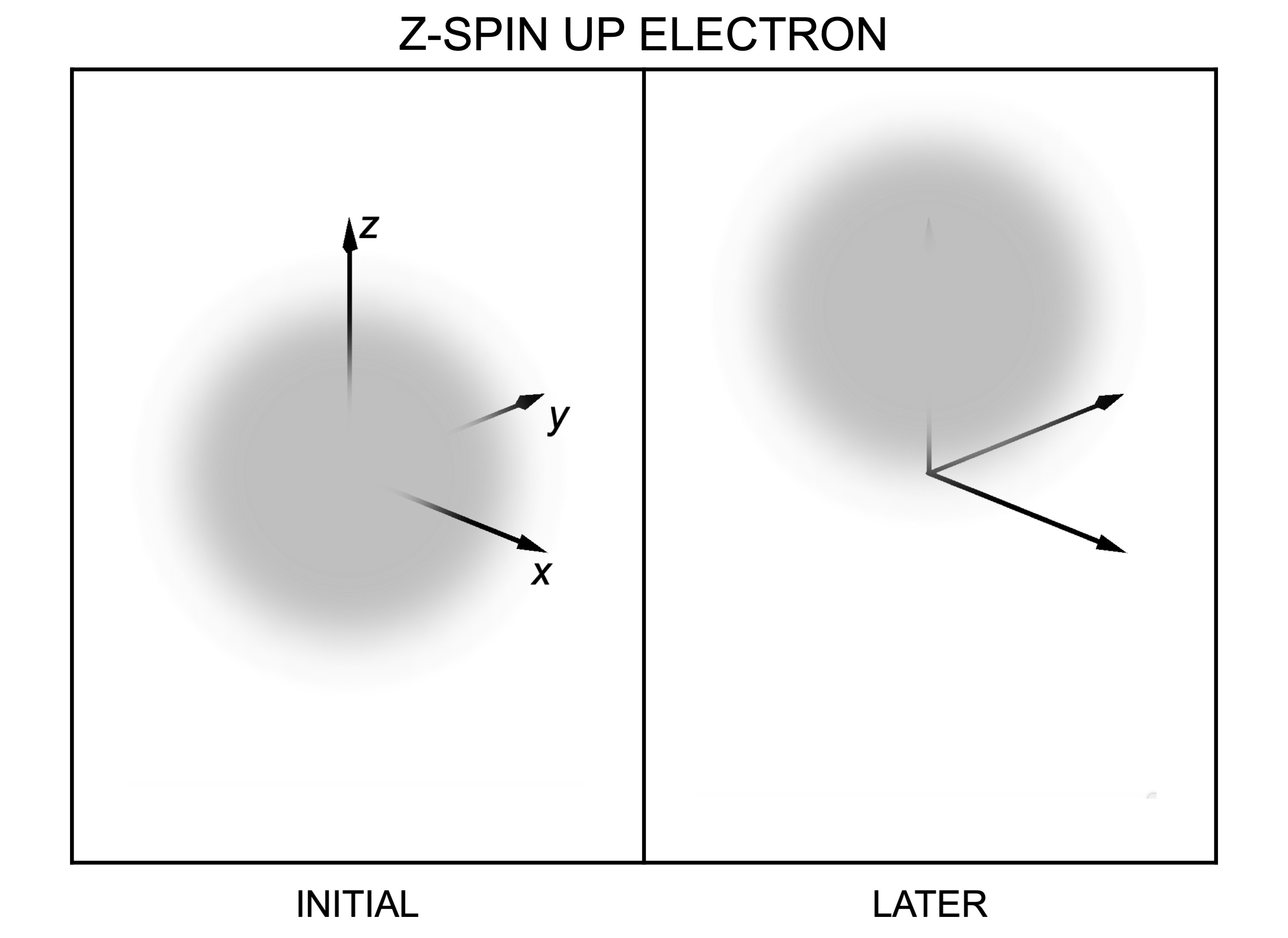}}\\
\center{\includegraphics[width=12cm]{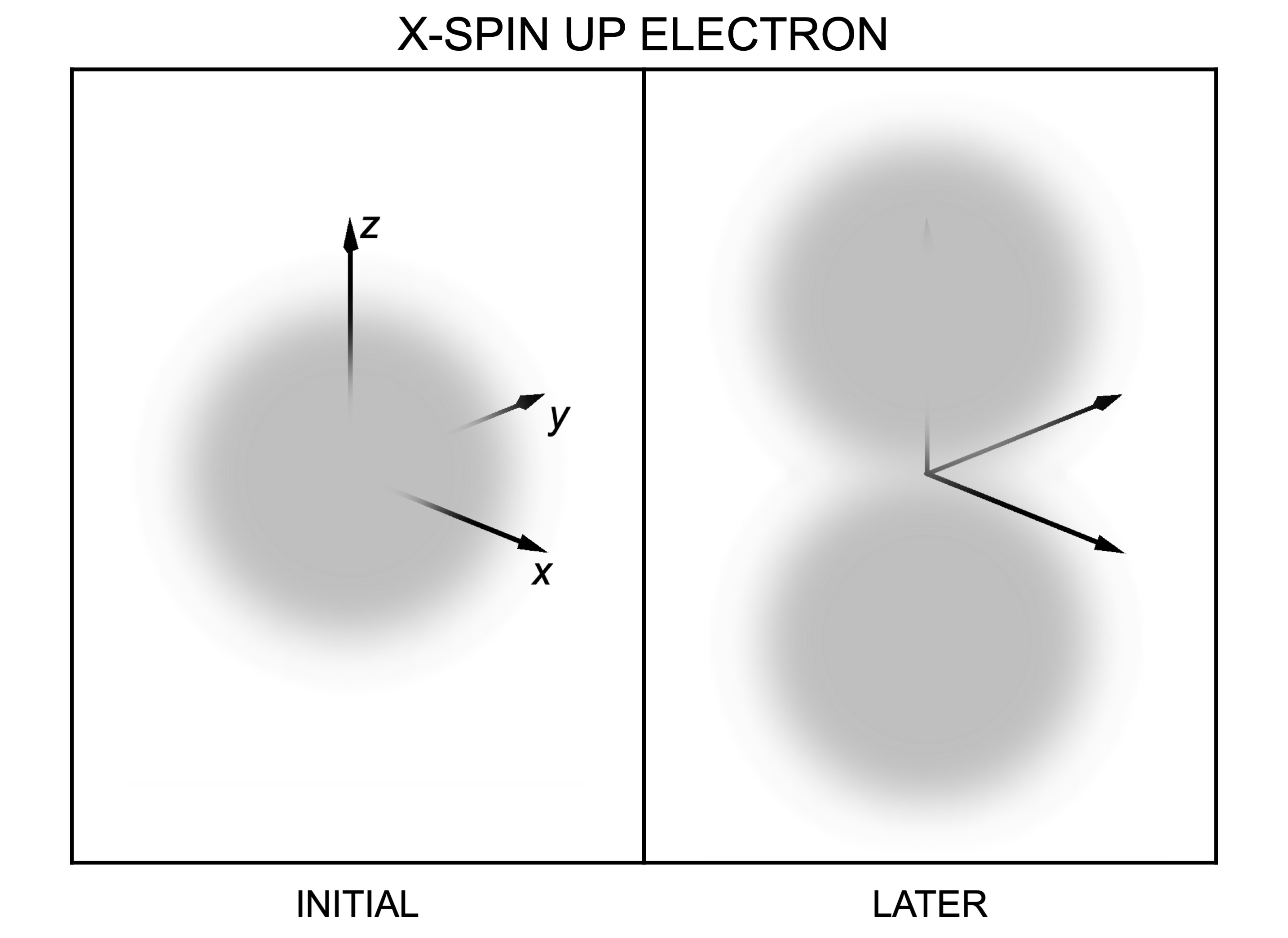}}
\caption{The first plot shows the probability density $\chi^{\dagger}\chi$ for the initial $z$-spin up wave function in \eqref{zupNRQM}, representing this density as a gray cloud of varying opacity.  The second plot shows the probability density at a later time, once the electron has passed through the magnetic field and its wave function has begun moving upwards.  The third plot shows the probability density for the $x$-spin up wave function in \eqref{xupNRQM}, which is identical to the probability density in the first plot.  The fourth plot shows the probability density at a later time, when the electron's wave function has split into two separate wave packets (one moving upwards and the other moving downwards).  Further embellishments (like the probability flux density) could be added to these plots, but here they are omitted to keep things simple.}
  \label{SG2}
\end{figure}

The future time evolution of the wave function after passing through the magnetic field \eqref{zupNRQMafter} can be solved exactly,\footnote{Similar examples of time evolution for gaussian wave packets appear in \citet[problem 2.43]{griffithsQM}; \citet[problem 5.8]{ballentine}.} though the result is somewhat complicated as it includes both the movement of the wave packet as a whole and the spreading of the wave packet,
\begin{align}
\chi(\vec{x},t) &= \left(\frac{1}{1+ \frac{i \hbar t}{m d^2}}\right)^{3/2} \left(\frac{1}{\pi d^2}\right)^{3/4} 
\\
&\quad\times \exp\left[\frac{-\frac{1}{2d^2}|\vec{x}-\frac{\mu \eta \Delta t}{m}t\hat{z}|^2 + \frac{i}{\hbar}\mu \eta z \Delta t+\left(\frac{i \hbar |\vec{x}|^2}{2 m d^4}-\frac{i \mu^2 \eta^2 \Delta t^2}{2 \hbar m}\right)t}{1+\frac{\hbar^2 t^2}{m^2 d^4}}- \frac{i}{\hbar}\mu B_0 \Delta t\right] \left(\begin{matrix} 1\\0 \end{matrix}\right)\ .
\end{align}
From the first term in the exponential, you can see that the center of the wave packet is moving upwards with a velocity of $\frac{\mu \eta \Delta t}{m} \hat{z}$, as in \eqref{zvelocitysphere}.  Figure \ref{SG2} shows the wave packet after some motion upwards but before significant spreading.

Let us now consider the behavior of an $x$-spin up electron in the Stern-Gerlach experiment.  We can take this electron's wave function to be an equal superposition of the $z$-spin up wave function in \eqref{zupNRQM} and a similar $z$-spin down wave function, yielding:
\begin{equation}
\chi(\vec{x}) = \left(\frac{1}{\pi d^2}\right)^{3/4}  \exp\left[\frac{-|\vec{x}|^2}{2 d^2}\right] \left(\begin{matrix} \frac{1}{\sqrt{2}}\\ \frac{1}{\sqrt{2}} \end{matrix}\right)
\ .
\label{xupNRQM}
\end{equation}
Evolving under \eqref{SGpauli} while in the magnetic field (again neglecting the $\mu \eta x \sigma_x$ term), the upper component of the wave function will pick up the same $z$-dependent phase factor as in \eqref{zupNRQMafter} and the lower component will pick up a similar contribution with the signs flipped,
\begin{equation}
\chi(\vec{x}) = \left(\frac{1}{\pi d^2}\right)^{3/4} \exp\left[\frac{-|\vec{x}|^2}{2 d^2} \right] \left(\begin{matrix} \frac{1}{\sqrt{2}}\exp\left[-\frac{i}{\hbar}\mu B_0 \Delta t + \frac{i}{\hbar}\mu \eta z \Delta t\right] \\ \frac{1}{\sqrt{2}}\exp\left[\frac{i}{\hbar}\mu B_0 \Delta t - \frac{i}{\hbar}\mu \eta z \Delta t\right] \end{matrix}\right)
\ .
\label{xupNRQMafter}
\end{equation}
The exponentials with opposite signs in the upper and lower components introduce a relative phase between these components that captures, at the quantum level, the Larmor precession of the electron's spin about the $z$ component of the Stern-Gerlach magnetic field.\footnote{This effect is discussed in \citet[example 4.3]{griffithsQM}; \citet[sec.\ 12.1]{ballentine}.  Larmor precession is often ignored in simplified and idealized presentations of Stern-Gerlach experiments.  This precession is particularly relevant in repeated series of Stern-Gerlach setups, such as two-path experiments.  For example, one often assumes that if an $x$-spin up electron passes through the Stern-Gerlach magnetic field and then the two wave packets that diverge are brought back together, a subsequent $x$-spin measurement will show that the electron is $x$-spin up (when in actuality the electron's spin has been rotated by the first magnetic field).  This kind of simplified treatment appears in \citet[pg.\ 7--11]{albertQM}; \citet[pg.\ 22--25]{maudlinQM}; \citet[ch.\ 2]{barrettQM}.}  The $\frac{i}{\hbar}\mu \eta z \Delta t$ terms in the exponentials alter the future time evolution of this wave function.  Although the expectation value of momentum for the total wave function is zero, the upper component considered alone (and normalized) has an expectation value of $\mu \eta \Delta t \hat{z}$ and the lower component has an expectation value of $-\mu \eta \Delta t \hat{z}$.  As the wave function evolves, these two parts will go their separate ways and we will end up with a quantum superposition of a $z$-spin up wave packet moving upwards and and a $z$-spin down wave packet moving downwards.

At the beginning, we set out two key features of the Stern-Gerlach experiment that needed to be accounted for:\ uniqueness and discreteness.  The classical treatments in the preceding two sections had unique outcomes but did not capture the discreteness of possible outcomes (allowing the electron to be deflected upwards, downwards, or anywhere in between).  Thus far, in non-relativistic quantum mechanics we have seen discreteness emerge but not yet uniqueness.  If we calculate time evolution using the Pauli equation, the $x$-spin up wave packet splits into two parts, one deflected upwards and the other deflected downwards.  Why is it that we only see one unique outcome, with equal probability of the electron being found in the upper or lower region?  This, in brief, is the measurement problem.  Versions of quantum mechanics that attempt to solve this problem are called ``interpretations of quantum mechanics,'' though that terminology downplays the differences between them.  Let us briefly review three of the leading options for solving the measurement problem.\footnote{For detailed introductions to these interpretations of quantum mechanics, see the references in footnote \ref{textbooks}.}

One way to ensure unique outcomes is to modify the time evolution of the wave function so that it does not always evolve under the Pauli equation.  Suppose that sometimes instead the wave function collapses.  In particular, when the electron's final location is measured at the end of the experiment, its wave function collapses so that it collects itself either in the upper region or the lower region, with probabilities for each of the possibilities found by integrating $\chi^{\dagger}\chi$ for each of the two wave packets.  For the wave function in \eqref{xupNRQMafter}, after the wave packets have been given time to separate, this would yield a fifty percent chance of collapse to the upper region and a fifty percent chance of collapse to the lower region.  To fully develop this sort of proposal, one needs to be precise about exactly when the Pauli equation is violated and how the wave function evolves in those situations.  Although one might attempt to do so by claiming collapses occur upon measurement and trying to figure out exactly what should count as a measurement, there are more promising strategies available---such as Ghirardi-Rimini-Weber (GRW) theory---where collapses are almost certain to occur when large numbers of quantum particles become entangled.

A second way to ensure unique outcomes is to say that the point electron of section \ref{pointparticlesection} should be retained in the quantum treatment, with its dynamics modified so that its path is altered by the presence of a quantum wave function obeying the Pauli equation.  Bohmian mechanics takes this route and includes a precise law, the guidance equation, specifying how the point electron moves.  In the $z$-spin up example above, the electron will be swept along with the wave packet and deflected upwards. In the $x$-spin up example, it might be swept along with the packet deflected upwards or the packet deflected downwards (depending on its exact initial position, which would not be known).  In different developments of Bohmian mechanics, the electron may or may not itself possess an intrinsic magnetic moment pointing in a particular direction.\footnote{\citet[ch.\ 7]{albertQM}; \citet[ch.\ 10]{bohmhiley}; \citet[sec.\ 8.4]{durrtteufel}; \citet[pg.\ 346]{norsen2014}, for example, do not attribute an intrinsic magnetic moment to the electron itself.}  If it does have such a property, one must add a law specifying how the magnetic moment evolves (\citealp{dewdney1986}; \citealp[ch.\ 9]{holland}).  When this is done for the $x$-spin up example, the electron would start with its magnetic moment aligned along the $x$ axis and the Stern-Gerlach experiment would then force the electron to align its magnetic moment along the $z$ axis (pointing either up or down, depending on which of the two wave packets the electron ends up in).

A third option is to deny that there really is a unique outcome and instead just try to explain why it appears to us that there is just one outcome.  According to the many-worlds interpretation, the $x$-spin up wave packet splits into two pieces:\ a $z$-spin up wave packet that gets deflected upwards and becomes entangled with a measuring device that shows up as the result, and a $z$-spin down wave packet that gets deflected downwards and becomes entangled with a measuring device that shows down as the result.  What we end up with is a quantum superposition of the two possible results, a superposition of entire worlds where each outcome happens.  In each world, the experiment has a unique result and the version of the experimenter in that world sees just that one result.  Thus, the many-worlds interpretation accounts for the apparent uniqueness of measurement outcomes.

We could continue listing options and then go on to compare the virtues of each proposal.  But, that would take us too far afield.  My point here is that although discreteness (two-valuedness) can be derived straightforwardly from non-relativistic quantum mechanics, the correct explanation for uniqueness is controversial.  Still, there are a variety of options and, in the end, physicists and philosophers agree that somehow non-relativistic quantum mechanics can account for both the discreteness and the (at least apparent) uniqueness of outcomes in the Stern-Gerlach experiment.  When I say that non-relativistic quantum mechanics can explain these features of Stern-Gerlach experiments, I mean that the theory can do so once it has been formulated in a precise way and the measurement problem has been solved.

Before moving on, it is worth noting that none of the options canvassed above would judge the Stern-Gerlach experiment to be a true measurement of some pre-existing fact about whether the electron was initially $z$-spin up or $z$-spin down.\footnote{This point is explained clearly in relation to collapse theories by \citet[pg.\ 101--102]{maudlinQM} and in relation to Bohmian mechanics by \citet{norsen2014}.}  Thus, the name ``Stern-Gerlach experiment'' is better than ``spin measurement.''  However, I will at times use both because the terminology of ``measurement'' is, at this point, deeply entrenched.

\section{Relativistic Quantum Mechanics}\label{rqmsection}

Although we are not going to imagine sending electrons through the Stern-Gerlach experiment at relativistic speeds, it will be valuable to see how the experiment is treated using the tools of relativistic quantum mechanics (applied in the non-relativistic limit) because the classical field analysis in the next section will use the same mathematics.  In the context of relativistic quantum mechanics, the electron is modeled by a four-component wave function $\psi$ that evolves via the Dirac equation, including an external classical electromagnetic field described by the vector potential $\vec{A}$ and the scalar potential $\phi$,
\begin{equation}
i \hbar \frac{\partial \psi}{\partial t}=\big(\underbrace{-i \hbar c \: \vec{\alpha}\cdot\vec{\nabla} + \beta m c^2}_{\widehat{H}_0} + \underbrace{e\: \vec{\alpha} \cdot \vec{A} - e \phi}_{\widehat{H}_I} \big)\psi
\ .
\label{dirac}
\end{equation}
As in \eqref{pauli}, the full Hamiltonian is divided into a free term and an interaction term.\footnote{\citet[pg.\ 11]{bjorkendrell} explain the interaction term as an operator version of the classical potential energy of a point charge in a static electromagnetic field \citep[sec.\ 15.6]{feynman2}.  This raises a number of questions, as we noted earlier (in section \ref{rigidbodysection}) that potential energies are insufficient for achieving conservation of energy in classical electromagnetism and, putting that aside, the additional potential energy associated with the electron's intrinsic magnetic moment \eqref{penergy} has not been included.}  The alpha and beta matrices that appear in this equation can be written in terms of the Pauli matrices \eqref{matrixdefs} and the $2 \times 2$ identity matrix $I$ as
\begin{equation}
\vec{\alpha}=\left(\begin{matrix} 0&\vec{\sigma}\\ \vec{\sigma}&0 \end{matrix}\right) \quad\quad \beta = \left(\begin{matrix} I&0\\ 0&-I \end{matrix}\right)
\ .
\label{matrixdefs2}
\end{equation}

Because we are focusing on the non-relativistic limit of relativistic quantum mechanics, we will be able to reuse our wave functions from the previous section in analyzing the Stern-Gerlach experiment.  To prepare for this repurposing, let us review the derivation of the Pauli equation as a non-relativistic approximation to the Dirac equation.\footnote{See \citet[sec.\ 1.4]{bjorkendrell}; \citet[sec.\ 33]{lifshitzRQM}; \citet[sec.\ 10.4]{bohmhiley}; \citet[sec.\ 2.6]{ryder}; \citet{nowakowski1999}.}  In the non-relativistic limit, the time evolution of $\psi$ via the Dirac equation is dominated by the rest energy term, $\beta m c^2$, in the Hamiltonian.  Focusing on the positive rest energy $m c^2$, we can separate out the fast time evolution associated with this energy by writing the wave function as
\begin{equation}
\psi= \exp\left[-\frac{i}{\hbar} m c^2 t \right] \left(\begin{matrix} \chi_u \\ \chi_l \end{matrix}\right)
\ ,
\label{psidecomposition}
\end{equation}
where here we distinguish two two-component pieces of $\psi$:\ an upper piece $\chi_u$ and a lower piece $\chi_l$.  The Pauli equation will emerge as a non-relativistic description of the time evolution of $\chi_u$.  Inserting \eqref{psidecomposition} into \eqref{dirac} and using \eqref{matrixdefs2}, the Dirac equation becomes
\begin{equation}
i \hbar \frac{\partial}{\partial t} \left(\begin{matrix} \chi_u \\ \chi_l \end{matrix}\right) =-i \hbar c \left(\begin{matrix} \vec{\sigma}\cdot\vec{\nabla} \chi_l \\ \vec{\sigma}\cdot\vec{\nabla}\chi_u \end{matrix}\right) - 2 m c^2 \left(\begin{matrix} 0 \\ \chi_l \end{matrix}\right) + e\left(\begin{matrix} \vec{\sigma}\cdot\vec{A}\: \chi_l \\ \vec{\sigma}\cdot\vec{A}\: \chi_u \end{matrix}\right)  - e \phi \left(\begin{matrix} \chi_u \\ \chi_l \end{matrix}\right)
\ .
\label{brokenupDirac}
\end{equation}
If we assume that $\chi_l$ is varying slowly,\footnote{This assumption focuses our attention on positive energy modes, allowing us to set aside questions regarding the interpretation of negative energy modes (sometimes addressed by introducing an infinite ``sea'' of negative energy electrons---the Dirac sea).} we can use the lower part of \eqref{brokenupDirac} to write an approximate expression for $\chi_l$ in terms of $\chi_u$ as
\begin{equation}
\chi_l = \left( \frac{- i \hbar c \: \vec{\sigma}\cdot\vec{\nabla} + e \: \vec{\sigma}\cdot\vec{A}-e\phi I}{2 m c^2}\right)\chi_u
\ .
\label{upperlower}
\end{equation}
Plugging this expression for $\chi_l$ into the upper part of \eqref{brokenupDirac}, focusing on the Stern-Gerlach magnetic field as described by the vector potential in \eqref{SGvector}, and setting\footnote{By setting $\phi$ equal to zero we are choosing to ignore the electric field that would be present in the frame where the electron begins at rest (a choice that was explained in section \ref{rigidbodysection}).} $\phi=0$ yields (after some manipulation):
\begin{equation}
i \hbar \frac{\partial \chi_u}{\partial t} =\left( \frac{-\hbar^2}{2 m} \nabla^2 + \frac{e}{2 m c^2} |\vec{A}_{SG}|^2 + \mu\: \vec{\sigma} \cdot \vec{B}_{SG} \right) \chi_u
\ .
\label{almostpauli}
\end{equation}
Dropping the $|\vec{A}|^2$ term because it is suppressed by a factor of $\frac{1}{c^2}$, we see that $\chi_u$ obeys the Pauli equation \eqref{pauli},
\begin{equation}
i \hbar \frac{\partial \chi_u}{\partial t} = \Big(\frac{- \hbar^2}{2 m}\nabla^2+\mu\: \vec{\sigma} \cdot \vec{B}_{SG} \Big)\chi_u
\ .
\label{pauli2}
\end{equation}

Let us now consider the evolution of a $z$-spin up wave function in the context of relativistic quantum mechanics.  The two-component wave function from \eqref{zupNRQM} can be straightforwardly extended to a four-component wave function by treating it as $\chi_u$ and using \eqref{upperlower} to find $\chi_l$,\footnote{The state in \eqref{zupRQM} also appears in \citet[eq.\ 36]{howelectronsspin}.  Note that by taking a non-relativistic approximation we have ended up with a wave function that is not normalized.}
\begin{equation}
\psi(\vec{x}) = \left(\frac{1}{\pi d^2}\right)^{3/4}  \exp\left[\frac{-|\vec{x}|^2}{2 d^2}\right] \left(\begin{matrix} 1\\0\\ \frac{\hbar}{2 m c d^2}(i z) \\ \frac{\hbar}{2 m c d^2}(i x-y)\end{matrix}\right)
\ ,
\label{zupRQM}
\end{equation}
where here we assume that the electron has not yet entered the magnetic field of the Stern-Gerlach experiment.  As we have just shown that the Pauli equation can be used to approximate the time evolution of $\chi_u$, we can write the state immediately after the electron has left the magnetic field using the result from \eqref{zupNRQMafter} along with \eqref{psidecomposition} and \eqref{upperlower},
\begin{equation}
\psi(\vec{x}) = \left(\frac{1}{\pi d^2}\right)^{3/4}  \exp\left[\frac{-|\vec{x}|^2}{2 d^2} - \frac{i}{\hbar}\mu B_0 \Delta t + \frac{i}{\hbar}\mu \eta z \Delta t -\frac{i}{\hbar} m c^2 \Delta t\right] \left(\begin{matrix} 1\\0\\ \frac{\hbar}{2 m c d^2}(i z) + \frac{\mu \eta}{2 m c} \Delta t \\ \frac{\hbar}{2 m c d^2}(i x-y) \end{matrix}\right)
\ .
\label{zupRQMafter}
\end{equation}
As for \eqref{zupNRQMafter}, the $\frac{i}{\hbar}\mu \eta z \Delta t$ term in the exponential will make this wave packet move upwards as it evolves by the free Dirac equation.

We can repeat this procedure of porting results from the previous section for an $x$-spin up electron that passes through the Stern-Gerlach experiment.  The initial state would be
\begin{equation}
\psi(\vec{x}) = \left(\frac{1}{\pi d^2}\right)^{3/4}  \exp\left[\frac{-|\vec{x}|^2}{2 d^2}\right] \left(\begin{matrix} \frac{1}{\sqrt{2}}\\\frac{1}{\sqrt{2}}\\ \frac{1}{\sqrt{2}}\left(\frac{\hbar}{2 m c d^2}(ix+y+iz)\right) \\ \frac{1}{\sqrt{2}} \left(\frac{\hbar}{2 m c d^2}(i x-y-iz)\right)\end{matrix}\right)
\ ,
\label{xupRQM}
\end{equation}
and the state after passing through the magnets would be
\begin{align}
\psi(\vec{x}) &= \left(\frac{1}{\pi d^2}\right)^{3/4}  \exp\left[\frac{-|\vec{x}|^2}{2 d^2} -\frac{i}{\hbar} m c^2 \Delta t\right]
\nonumber
\\
&\quad\quad\times\left\{
 \frac{1}{\sqrt{2}}\exp\left[-\frac{i}{\hbar}\mu B_0 \Delta t + \frac{i}{\hbar}\mu \eta z \Delta t\right]  \left(\begin{matrix}
 1 \\
 0 \\
 \frac{\hbar}{2 m c d^2}(iz) + \frac{\mu \eta}{2 m c} \Delta t \\
 \frac{\hbar}{2 m c d^2}(i x-y)
 \end{matrix}\right)\right.
 \nonumber
\\
&\quad\quad\quad\quad+
\left.
 \frac{1}{\sqrt{2}}\exp\left[\frac{i}{\hbar}\mu B_0 \Delta t - \frac{i}{\hbar}\mu \eta z \Delta t\right]  \left(\begin{matrix}
 0 \\
 1 \\
 \frac{\hbar}{2 m c d^2}(ix+y)  \\
 \frac{\hbar}{2 m c d^2}(-iz) + \frac{\mu \eta}{2 m c} \Delta t
 \end{matrix}\right)
\right\}
\ .
\label{xupRQMafter}
\end{align}
As for \eqref{xupNRQMafter}, the $\frac{i}{\hbar}\mu \eta z \Delta t$ terms in the exponentials will cause this wave packet to split into a $z$-spin up wave packet moving upwards, like \eqref{zupRQMafter}, and a $z$-spin down wave packet moving downwards.

We have just seen how the Dirac equation delivers a discrete set of two possible outcomes (the particle being deflected upwards or downwards).  As in the previous section, to explain why the Stern-Gerlach experiment has a unique outcome we must adopt some solution to the measurement problem.  The three strategies discussed earlier can all be applied to the relativistic quantum mechanics of an electron obeying the Dirac equation.  The fact that the many-worlds interpretation can be extended straightforwardly to more complicated quantum theories, like relativistic quantum mechanics, has been touted as a central virtue of the interpretation (\citealp[sec.\ 1.7]{wallaceQM}; \citealp{wallace2020}).  Bohmian mechanics can also be extended to relativistic quantum mechanics, though for multiple particles this involves a privileged foliation and thus what you end up with is arguably not a truly relativistic theory (\citealp{bohm1953}; \citealp[ch.\ 12]{bohmhiley}; \citealp[sec.\ 12.2]{holland}; \citealp{durr2014}; \citealp[sec.\ 3.1]{tumulka2018}).  Extending GRW is possible as well (\citealp{tumulka2006}; \citealp{bedingham2014}; \citeauthor{maudlin2011}, \citeyear{maudlin2011}, ch.\ 10, \citeyear{maudlinQM}, ch.\ 7).

\section{Classical Field Theory}\label{cftsection}

Up to this point, we have reviewed the standard story about Stern-Gerlach experiments in some detail.  We have seen that both the uniqueness and discreteness (two-valuedness) of outcomes can be explained in non-relativistic and relativistic quantum theories that describe a point electron in a quantum superposition of different states.  In contrast, we could explain uniqueness but not discreteness in the classical theories that modeled the electron as either a rigid sphere or a point particle with intrinsic magnetic moment and angular momentum.  Thus, one might conclude that discreteness of outcomes is a distinctively quantum feature that cannot be captured in a classical theory.  This lesson would fit with Wolfgang Pauli's famous early description of spin as a ``classically non-describable two-valuedness.''\footnote{See \citet{pauli1925, pauli1946}; \citet[ch.\ 2]{tomonaga1997}; \citet{morrison2007, giulini2008}; \citet[sec.\ 7.4]{deregt}.}

I would now like to challenge that lesson by analyzing and advocating a third classical model of the electron that describes the two-valuedness (the discreteness) but not the uniqueness of outcomes (uniqueness becoming a quantum feature of spin).  In this classical model, the electron is treated as a lump of energy and charge in the classical Dirac field.  Mathematically, the electron is treated exactly as in the relativistic quantum mechanics of section \ref{rqmsection}.  However, the physical interpretation of this mathematics is different.  In section \ref{rqmsection}, $\psi$ was treated as a four-component complex-valued quantum wave function.  In this section, $\psi$ is treated as a four-component complex-valued classical field.  As we will see shortly, this classical field description of the electron fits within the framework of relativistic continuum mechanics, including currents and forces that closely resemble those of the rigid body analysis in section \ref{rigidbodysection}.  The extensive groundwork that we have laid in the previous sections, particularly \ref{rigidbodysection} and \ref{rqmsection}, will allow us to move quickly through this section's novel treatment of the Stern-Gerlach experiment.

In this classical context, we can take the Dirac field to obey the Dirac equation \eqref{dirac} and the electromagnetic field to obey Maxwell's equations, with the charge and current densities of the Dirac field acting as source terms.  These coupled relativistic equations are sometimes called the ``Maxwell-Dirac equations.''\footnote{See, for example, \citet{gross1966, glassey1979, flato1987}.}  For our purposes, we can use the following simple expressions for the charge and current densities of the Dirac field,\footnote{I have argued elsewhere that we should modify equations like \eqref{diracchargedensity} and \eqref{diraccurrentdensity} so that the negative frequency modes of the Dirac field are associated with negative charge and positive energy \citep{positrons}.  However, we don't have to be worry about that here because the negative frequency modes are not important in the non-relativistic approximation that we are using \citep[pg.\ 10]{bjorkendrell}.}
\begin{align}
\rho^q&=-e \psi^\dagger \psi
\label{diracchargedensity}
\\
\vec{J}&=-e c \psi^\dagger \vec{\alpha} \psi
\label{diraccurrentdensity}\ .
\end{align}
In the relativistic quantum mechanics of section \ref{rqmsection}, $\psi^\dagger \psi$ is interpreted as a probability density and $c \psi^\dagger \vec{\alpha} \psi$ as a probability flux density.  In the classical context of this section, we jettison that interpretation and instead multiply those quantities by $-e$ and view them as the charge density and charge flux density (current density) of the Dirac field.

As the classical electromagnetic and Dirac fields interact with one another, they exchange momentum and energy.  We can interpret the rate at which momentum is transferred from the electromagnetic field to the Dirac field per unit volume as a density of force exerted by the electromagnetic field on the Dirac field.\footnote{For an examination and defense of the idea that forces can be exerted upon fields, see \citet{forcesonfields}.}  From Maxwell's equations and the expressions for the momentum density and momentum flux density of the electromagnetic field, one can show\footnote{Consider, for example, working backwards through the proofs in \citet[sec.\ 6.7]{jackson}; \citet[sec.\ 8.2]{griffiths}.} that the rate at which momentum is transferred per unit volume is given by the Lorentz force law \eqref{cforcelaw}.  So, we can use the Lorentz force law to calculate the forces exerted by the electromagnetic field on an electron even in this unfamiliar context where it is being modeled as part of a classical field.

To fully describe the interactions between an electron and the electromagnetic field, we would need to consider the electron's own contribution to the electromagnetic field and how the electron reacts to that contribution.\footnote{For philosophical discussion of self-interaction, see \citet{lange, frisch2005, earman2011, lazarovici2018, maudlin2018, hartensteinhubert}.}  This complicates the analysis and introduces a problem of self-repulsion:\ the electron would produce a very strong inwardly directed electric field that would exert strong outward forces on all of the parts of the electron.  In the absence of anything holding the electron together,\footnote{Hypothetical forces holding the electron together have been called ``Poincar\'{e} stresses.''  See \citet[ch.\ 28]{feynman2}; \citet{rohrlich1973}; \citet{pearle1982}; \citet{schwinger1983electromagnetic}; \citet[ch.\ 16]{jackson}; \citet[sec.\ 6.3]{rohrlich}; \citet[sec.\ 5]{griffithsletter}.} these forces would cause the electron to rapidly explode.  This is a serious problem and not one I intend to solve here.  Let us put the electron's contribution to the electromagnetic field aside\footnote{The electron's contribution to the electromagnetic field is also relevant for a precise calculation of the total angular momentum of the electron itself and the electromagnetic field that surrounds it, as there is angular momentum in both the Dirac and electromagnetic fields \citep{howelectronsspin}.} and treat the electron as reacting only to the external magnetic field \eqref{SGfield}.

\begin{figure}[p!]
\center{\includegraphics[width=12cm]{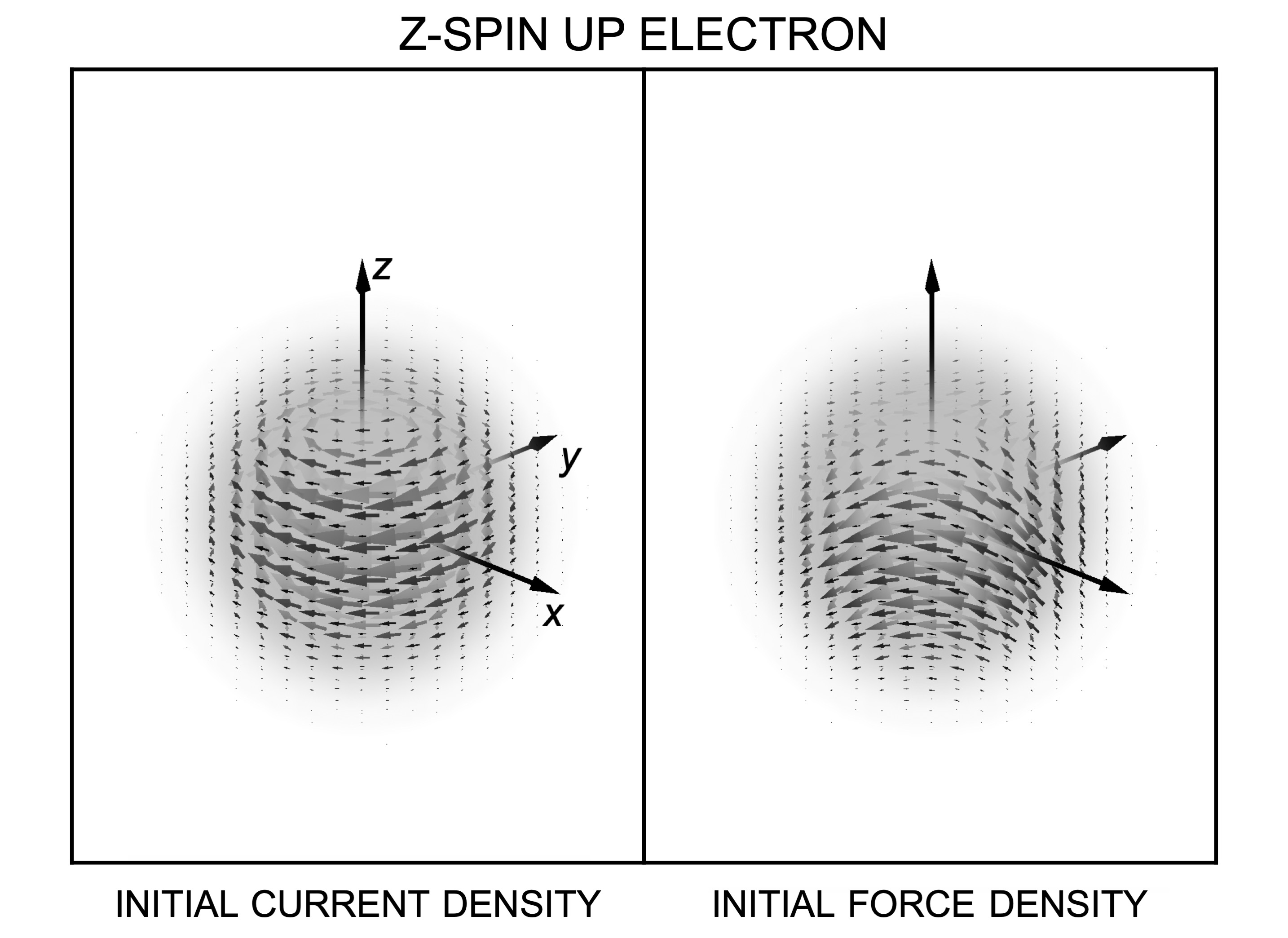}}\\
\vspace*{-4pt}
\center{\includegraphics[width=12cm]{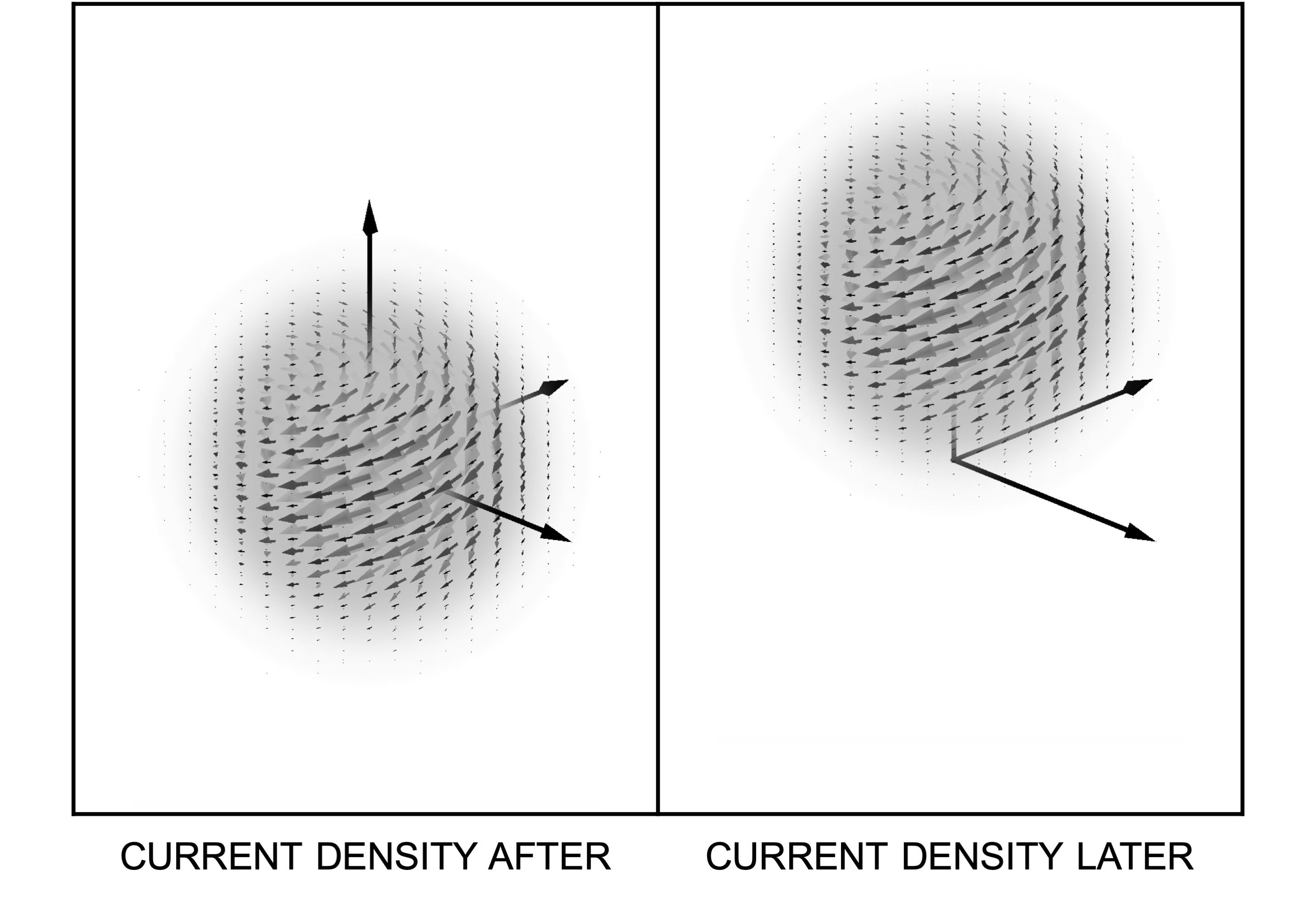}}
\caption{The first plot depicts a $z$-spin up electron in classical Dirac field theory, including a gray cloud representing the electron's charge density \eqref{zupchargeblob} and arrows depicting the flow of charge around the $z$ axis \eqref{zupcurrentblob}.  The second plot shows the density of force \eqref{zupforcedensityblob} on that electron when it enters the inhomogeneous magnetic field of the Stern-Gerlach apparatus.  As in figure \ref{SG1}, the variation of the magnetic field across the diameter of the electron is exaggerated so that you can easily see the upward force on the electron.  The third plot illustrates how the current density is altered \eqref{zupcurrentblobafter} by the electron spending time in the inhomogeneous magnetic field, picking up a downwards component corresponding to the upwards motion of the electron.  The fourth plot shows that electron at a later time, once it has moved a small distance upwards.}
  \label{SG3}
\end{figure}

With that stage setting complete, we will now consider the behavior of $z$-spin up and $x$-spin up electrons in the Stern-Gerlach experiment (as we have done in each of the previous sections).  For a $z$-spin up electron, we can use the state from \eqref{zupRQM} as an initial description of the classical Dirac field where the electron's charge is distributed over a gaussian wave packet centered at the origin, with a charge density of approximately
\begin{equation}
\rho^q=-e\left(\frac{1}{\pi d^2}\right)^{3/2}\exp\left[\frac{-|\vec{x}|^2}{d^2}\right] 
\ ,
\label{zupchargeblob}
\end{equation}
assuming $d\gg\frac{\hbar}{mc}$.\footnote{This assumption is part of our non-relativistic approximation (\citealp[pg.\ 39]{bjorkendrell}; \citealp[sec.\ 5]{howelectronsspin}).}  The current density for this state, calculated from \eqref{zupRQM} via \eqref{diraccurrentdensity}, is\footnote{This current density is discussed in \citet[sec.\ 4]{ohanian}; \citet[eq.\ 31]{howelectronsspin}.}
\begin{equation}
\vec{J}=\left(\frac{1}{\pi d^2}\right)^{3/2} \exp\left[\frac{-|\vec{x}|^2}{d^2}\right] \frac{2 \mu c}{d^2} (\vec{x} \times \hat{z})
\ ,
\label{zupcurrentblob}
\end{equation}
which is very similar to the earlier current density for a spherical electron \eqref{zupcurrentsphere} except that the current in \eqref{zupcurrentblob} becomes weaker (because the density of charge decreases) as you move outward.  Plugging this current into the Lorentz force law \eqref{cforcelaw}, we can calculate the density of force on the electron as it passes through the Stern-Gerlach magnetic field \eqref{SGfield},
\begin{align}
\vec{f}&=\frac{1}{c} \vec{J} \times \vec{B}_{SG}
\nonumber
\\
&=\frac{2 \mu}{d^2}\left(\frac{1}{\pi d^2}\right)^{3/2} \exp\left[\frac{-|\vec{x}|^2}{d^2}\right] \left( (-B_0 x + \eta x z) \hat{x} + (B_0 y - \eta y z)\hat{y} + \eta x^2 \hat{z} 
\right)
\ ,
\label{zupforcedensityblob}
\end{align}
similar to \eqref{zupsphereforcedensity}.  As in section \ref{rigidbodysection}, it is the $x$ component of the magnetic field that is responsible for the net upward force on the electron and that force is concentrated on the electron's sides (pulling it up by its ears, as shown in figure \ref{SG3}).  Integrating the relevant term over all of space to calculate the total force on the electron yields
\begin{align}
\vec{F} &= \frac{2 \mu}{d^2}\left(\frac{1}{\pi d^2}\right)^{3/2}\int{d^3x\  \exp\left[\frac{-|\vec{x}|^2}{d^2}\right]\eta x^2} \hat{z}
\nonumber
\\
&= \mu \eta \hat{z}
\ ,
\end{align}
in agreement with the earlier calculations of the total force on a spherical electron \eqref{zforcesphere} or a point electron \eqref{zforcepoint}.

Because the dynamics are given by the Dirac equation, we can use our previous result for the state of the electron after passing through the magnetic field \eqref{zupRQMafter}.  Using \eqref{diraccurrentdensity}, we see that the current density,
\begin{equation}
\vec{J}=\left(\frac{1}{\pi d^2}\right)^{3/2} \exp\left[\frac{-|\vec{x}|^2}{d^2}\right]\left( \frac{2\mu c}{d^2} (\vec{x} \times \hat{z}) - \frac{e \mu \eta}{m}\Delta t \hat{z} \right)
\ ,
\label{zupcurrentblobafter}
\end{equation}
has picked up an additional term pointing downward and representing the upward motion of the wave packet (the signs for current and overall motion being opposite because the electron is negatively charged).  We can determine the newly acquired upward velocity by dividing this new term in the current density \eqref{zupcurrentblobafter} by the charge density \eqref{zupchargeblob}.  This gives a velocity of $\frac{\mu \eta \Delta t}{m} \hat{z}$, as in \eqref{zvelocitysphere}.

Next, let us consider sending an $x$-spin up electron through the Stern-Gerlach experiment.  We can take the electron's initial state to be given by \eqref{xupRQM} and use \eqref{diraccurrentdensity} to find its current density,
\begin{equation}
\vec{J}=\left(\frac{1}{\pi d^2}\right)^{3/2} \exp\left[\frac{-|\vec{x}|^2}{d^2}\right] \frac{2 \mu c}{d^2} (\vec{x} \times \hat{x})
\ ,
\label{xupcurrentblob}
\end{equation}
similar to \eqref{xupcurrentsphere}.  The Lorentz force law \eqref{cforcelaw} can be applied to calculate the density of force,
\begin{equation}
\vec{f}=\frac{2\mu}{d^2}\left(\frac{1}{\pi d^2}\right)^{3/2} \exp\left[\frac{-|\vec{x}|^2}{d^2}\right] \left( (B_0 z - \eta z^2) \hat{x} - \eta x y \hat{y} - \eta x z \hat{z} \right)
\ ,
\label{xupforcedensityblob}
\end{equation}
similar to \eqref{xupsphereforcedensity}.  As in section \ref{rigidbodysection}, the $- \eta z^2 \hat{x}$ term will not lead to a significant deflection of the electron because the direction of the net force on the electron will change rapidly as the electron undergoes Larmor precession.  We can see the effect of this Larmor precession by looking at the state of the electron after it has passed through the magnetic field, \eqref{xupRQMafter}.  The (somewhat complicated) current density for this state is
\begin{align}
\vec{J}=\left(\frac{1}{\pi d^2}\right)^{3/2} \exp\left[\frac{-|\vec{x}|^2}{d^2}\right]&\left\{ \left(\frac{2 \mu c}{d^2} (\vec{x} \times \hat{x}) - \frac{e \mu \eta}{m}\Delta t \hat{x} \right)\cos\left[\frac{2 \mu}{\hbar} (B_0 - \eta z)\Delta t\right] \right.
\nonumber
\\
&\quad\quad\left.+\left(\frac{2 \mu c}{d^2} (\vec{x} \times \hat{y}) - \frac{e \mu \eta}{m}\Delta t \hat{y} \right)\sin\left[\frac{2 \mu}{\hbar} (B_0 - \eta z)\Delta t\right]\right\}
\ .
\label{xupcurrentblobafter}
\end{align}
If the magnetic field were homogenous ($\eta=0$), the magnetic moment of the electron would be rotating (with the electron starting $x$-spin up) about the $z$ axis at a uniform angular frequency of $\frac{2\mu}{\hbar} B_0$ as we imagine longer durations $\Delta t$ (the same frequency of Larmor precession that was calculated at the end of section \ref{rigidbodysection}).  However, because of the inhomogeneity of the magnetic field ($\eta \neq 0$), the rate of Larmor precession is actually slower at the top of the wave packet and faster at the bottom.  Note that the current density in \eqref{xupcurrentblobafter} is divergenceless and thus at this moment the charge density is not changing.

\begin{figure}[p!]
\center{\includegraphics[width=12cm]{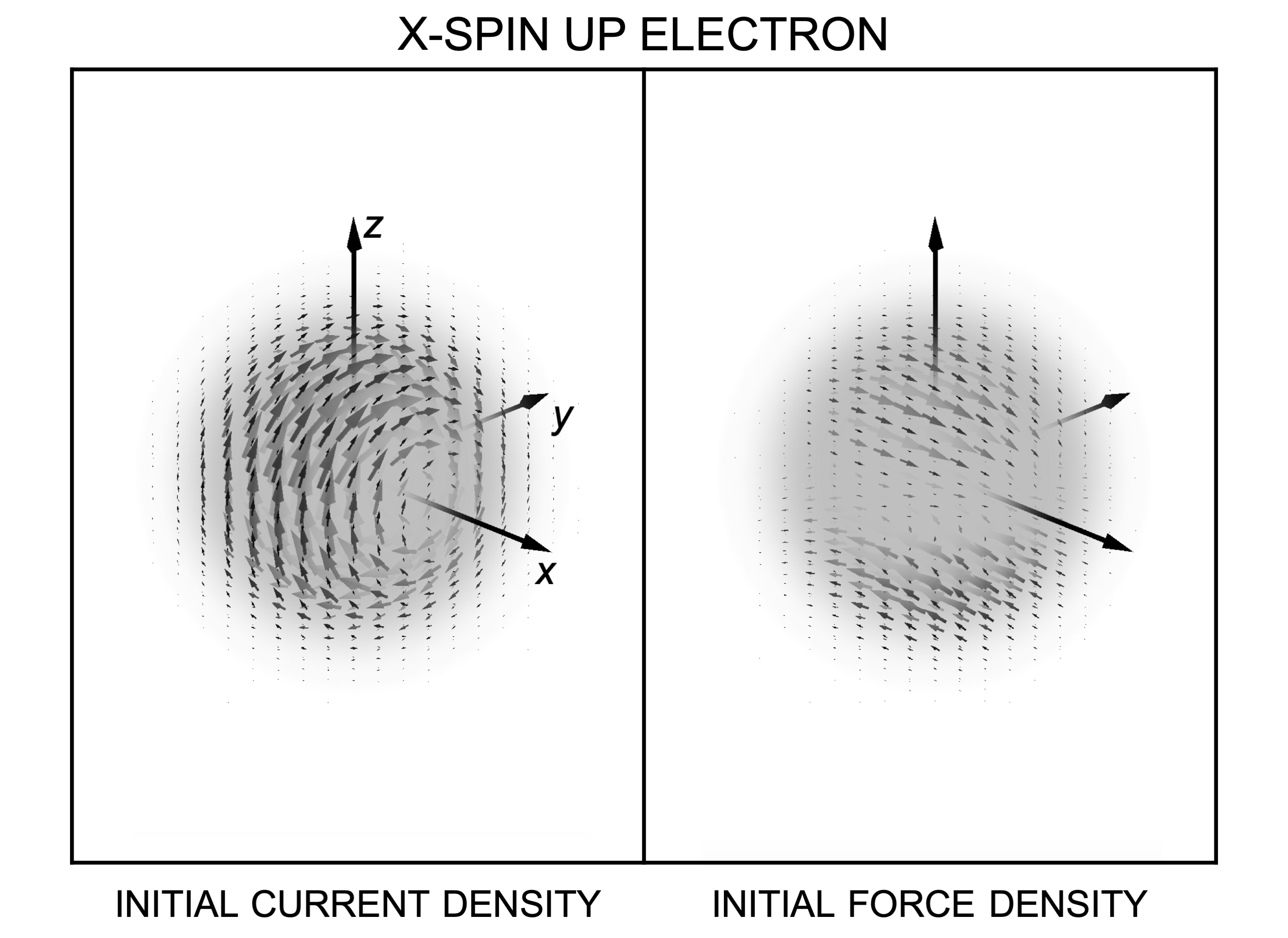}}\\
\vspace*{-4pt}
\center{\includegraphics[width=12cm]{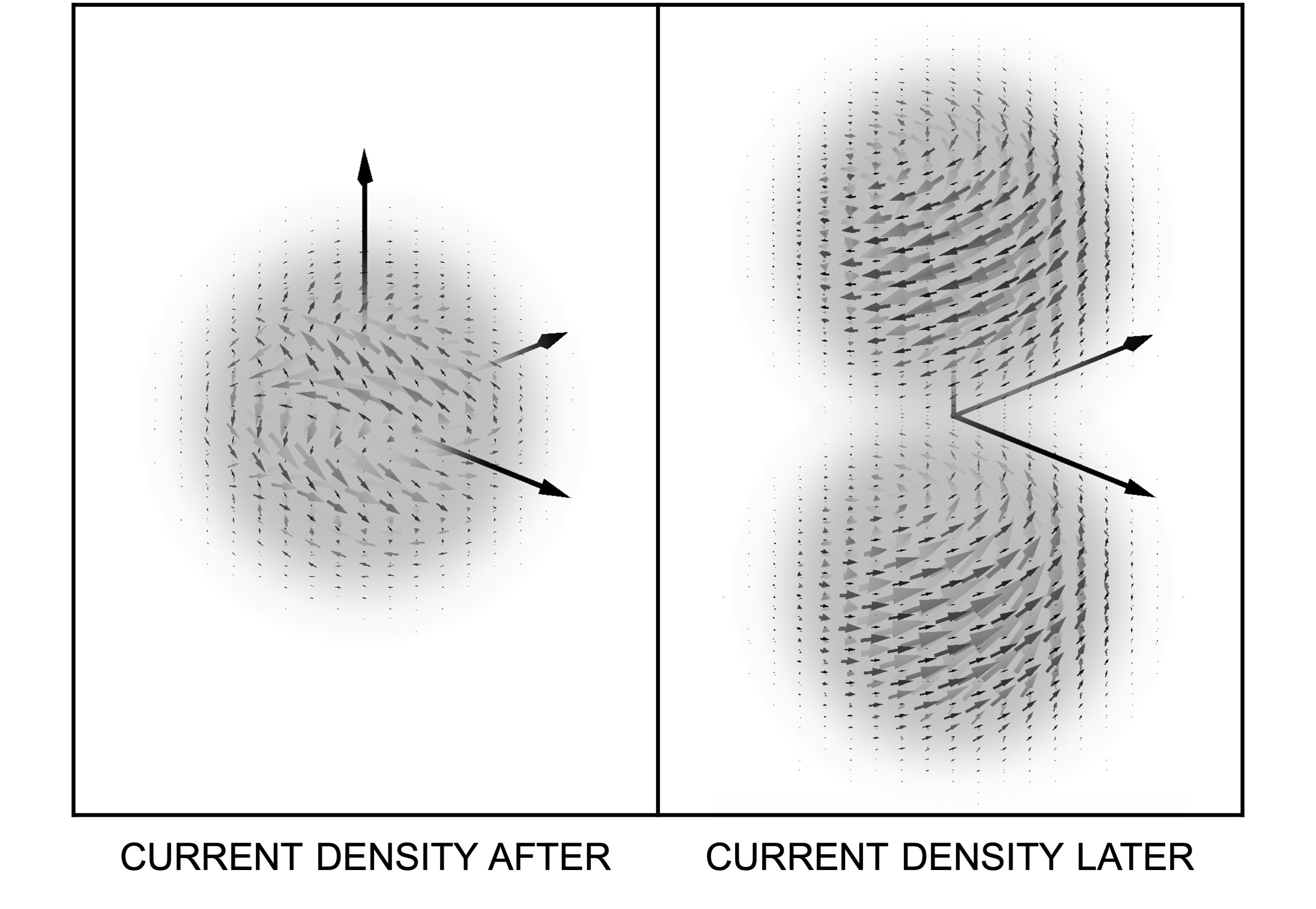}}
\caption{The first two plots show the density of charge, current, and force for an $x$-spin up electron entering the Stern-Gerlach magnetic field, \eqref{xupcurrentblob} and \eqref{xupforcedensityblob}.  The third plot depicts the jumbled current after the electron has spent some time in the magnetic field \eqref{xupcurrentblobafter}.  In the fourth plot, we see the electron at a later time when it has split into a $z$-spin up piece that is moving upwards and a $z$-spin down piece that is moving downwards.}
  \label{SG4}
\end{figure}

When we then let the state in \eqref{xupRQMafter} evolve under the free Dirac equation, it will split into two wave packets---with the current in the upper packet resembling \eqref{zupcurrentblob} and the current in the lower packet being oppositely oriented.  In the Stern-Gerlach experiment, a single electron initially spinning about the $x$ axis has its current jumbled and then splits into two separate halves, each of which is spinning about the $z$ axis (see figure \ref{SG4}).  As in the interpretations of quantum mechanics discussed at the end of section \ref{nrqmsection}, the Stern-Gerlach experiment does not act as a measurement of some preexisting $z$-spin in this classical context.  Instead, the experiment rips the electron into a $z$-spin up piece and $z$-spin down piece.  If initially the spin is pointing somewhere in the $x$-$y$ plane, these pieces will carry equal charge.  If it is pointing in some other direction, the pieces will carry unequal charge (there being only a single piece if the spin initially points along the $z$ axis).

Of course, this is not what we see when the experiment is conducted.  The classical field model of the electron explains the discreteness but not the uniqueness of outcomes.  To accurately account for the fact that we see the electron hit the detector at a single unique location when the Stern-Gerlach experiment is performed, we need to move from classical field theory to quantum field theory.  But, before we discuss that transition, let us take a moment to compare the classical models of the electron that we have studied.

In comparison to the classical rigid spherical electron of section \ref{rigidbodysection}, the classical Dirac field model of this section has a number of advantages.  First, recall that in the spherical electron model we had to postulate that the electron's magnetic moment and angular momentum have fixed values.  The mass and charge of the electron will never rotate faster or slower.  Accounting for the continued truth of this postulate would require modifying the usual laws of rigid body mechanics.  This particular complication does not arise in classical Dirac field theory, where the field just evolves by the Dirac equation.  Second, to find an adequate classical model of the electron we would need a relativistic theory and that would require abandoning true rigidity and positing some kind of force holding the electron together (a kind of force that, to be fair, may be necessary to overcome the problem of self-repulsion for the classical field model mentioned earlier in this section).  Third, and most important, none of the standard quantum theories that are used to describe the electron (in sections \ref{nrqmsection}, \ref{rqmsection}, and \ref{qftsection}) treat it as a spherical body in a quantum superposition of classical states.  Thus, for the purposes of interpreting our best extant quantum theories, the spherical model is of limited interest.  By contrast, if the approach to quantum field theory described in the next section is viable, then classical Dirac field theory can be taken as giving the classical states that quantum field theory describes as entering quantum superpositions.

The classical field model of the electron also has important advantages over the classical point particle model of section \ref{pointparticlesection}.  First, you do not need to add extra forces and torques to the Lorentz force law, as in \eqref{pforcelaw2} and \eqref{torquelaw}.  Instead, you start from simple field equations and derive the Lorentz force law as a consequence.  Second, the classical field theory of this section treats all angular momentum in a unified way:\ it always and only results from true rotation.\footnote{This section has focused on magnetic moment and the flow of charge.  Angular momentum results from the flow of relativistic mass, or you could say the flow of energy (as relativistic mass is proportional to energy).  This flow is analyzed in \citet{ohanian, howelectronsspin}.}  This is an improvement over the classical point particle mechanics of section \ref{pointparticlesection}, where we must view angular momentum as sometimes an intrinsic property of point particles and sometimes the result of actual rotation.  Third, in the classical field theory of this section the only sources for the electromagnetic field are charge densities and charge currents.  By contrast, the classical point particle theory of section \ref{pointparticlesection} must give a disunified account where electromagnetic fields are sourced by both the charges and the intrinsic magnetic moments of moving particles (each of which would have to appear in Maxwell's equations).\footnote{See footnote \ref{modifiedforcefootnote} for Griffiths' statement of this criticism.}  Fourth, the electromagnetic field sourced by a point electron would have infinite energy, as calculated from \eqref{EMenergydensity}, because the electric field becomes extremely strong as you approach the electron.  This problem of self-energy is related to the (also bad) problem of self-repulsion facing the classical field model of the electron, but is arguably more severe because of the infinities involved.

\section{Quantum Field Theory}\label{qftsection}

There is wide agreement that the results of the Stern-Gerlach experiment can be explained within quantum field theory, though this theory is more advanced than necessary (as compared to the quantum theories of sections \ref{nrqmsection} and \ref{rqmsection}) so it is not normal to analyze the Stern-Gerlach experiment in this context.  A detailed quantum field theoretic account of the experiment would be illuminating, but I will not provide one here.  Instead, I will only offer some brief remarks.

There is disagreement as to how quantum field theories should be formulated.  One locus of disagreement is the question of what classical entities quantum field theory describes as entering quantum superpositions.  On a particle approach to quantum field theory, it is point particles that enter quantum superpositions.  The quantum field theory of quantum electrodynamics is viewed as an extension of the relativistic quantum mechanics of section \ref{rqmsection} to handle multiple electrons, positrons, and photons (including the creation and annihilation of such particles).\footnote{Depending on the way the particle approach is executed, positrons may either be viewed as fundamental particles or as holes in the Dirac Sea.  Both options have been explored by scholars working to extend Bohmian mechanics to quantum field theory, the former by \citet{durr2004, durr2005} and the latter by Bohm himself (\citealp[pg.\ 275]{bohm1953}; \citealp[ch.\ 12]{bohmhiley}) and more recently by \citet{colin2007, deckert2019}.}  On a field approach, it is classical fields that enter quantum superpositions.  Quantum electrodynamics is arrived at by quantizing the classical theory of interacting Dirac and electromagnetic fields from section \ref{cftsection}.  If we adopt a particle approach to quantum field theory, then the standard story about Stern-Gerlach experiments is correct:\ the classical point particle theory that quantum field theory is built upon can explain the uniqueness of outcomes but not the discreteness (that is a distinctively quantum feature of spin).  If, on the other hand, we adopt a field approach then the story is different:\ the classical field theory that quantum field theory is built upon can explain the discreteness of outcomes in the Stern-Gerlach experiment but not the uniqueness (that is a distinctively quantum feature of spin).  Let us focus here on the prospects for modeling the Stern-Gerlach experiment within the field approach.\footnote{For more on the particle and field approaches to quantum field theory, see the references in \citet[footnote 2]{howelectronsspin}.}

In a field approach to the quantum field theory of quantum electrodynamics, the quantum state would be represented as a wave functional that assigns amplitudes to classical configurations of the Dirac and electromagnetic fields.  At the beginning of a Stern-Gerlach experiment, this wave functional might be peaked around a classical configuration of the Dirac field representing an electron that is $x$-spin up \eqref{xupRQM}.  At the end of the Stern-Gerlach experiment (perhaps only after interaction with the detector), the wave functional should be peaked around two classical configurations of the Dirac field, one representing an electron that is $z$-spin up (and has been deflected upwards) and another representing an electron that is $z$-spin down (and has been deflected upwards).  Then, as at the end of section \ref{nrqmsection}, we could employ an interpretation of quantum physics like GRW theory, Bohmian mechanics, or the many-worlds interpretation to explain why we observe a single unique outcome, either an electron that has been deflected upwards or an electron that has been deflected downwards, with equal probability for each of the two possible outcomes.

There are at least three challenges that must be overcome to complete the story in the previous paragraph.  The first challenge is that in the field approach to fermionic fields, like the Dirac field, it appears that the wave functional must assign amplitudes not to classical configurations of the complex-valued Dirac field from section \ref{cftsection}, but instead to a Dirac field whose values are anticommuting Grassmann numbers.  Further, it seems that the amplitudes assigned by the wave functional will themselves have to be Grassmann numbers.  This raises a number of questions.  In particular, this complication makes it difficult to understand how the amplitude squared of the wave functional could be interpreted as a probability density and how we can make sense of the above descriptions of wave functionals that are peaked around certain classical configurations.\footnote{Wave functionals over Grassmann-valued fields are used in \citet{floreanini1988, jackiw1990, hatfield, valentini1992, valentini1996}.  The issues mentioned above are discussed in \citet[sec.\ 9.2]{struyve2010}; \citet[sec.\ 3.3]{struyve2011}; \citet[appendix A]{positrons}.}

The second challenge is to extend existing interpretations of quantum physics to quantum field theory.\footnote{For discussion of the various difficulties involved in extending the interpretations of quantum mechanics from section \ref{nrqmsection} to relativistic quantum field theory, see \citet{wallace2008}; \citet{barrett2014}; \citet[ch.\ 7]{maudlinQM}; \citet[ch.\ 11--12]{durr2020}.}  This is an ongoing area of research and it may turn out that some promising interpretations of quantum mechanics cannot be extended into viable interpretations of quantum field theory.  The many-worlds interpretation appears to be in the strongest position here, as it is arguably trivial to extend the interpretation to quantum field theory on either the particle or field approach.\footnote{See \citet{wallace2008, wallace2018, wallace2020}; \citet[sec.\ 1.7]{wallaceQM}.}  The project of extending Bohmian mechanics to quantum field theory is well underway, sometimes pursued taking a particle approach and sometimes taking a field approach to quantum field theory.\footnote{See \citet{struyve2010, struyve2011, tumulka2018} and references therein.}  There has been less research on extending GRW theory, though the work cited at the end of section \ref{rqmsection} moves us closer to such an extension (taking a particle approach to quantum field theory).

The third challenge is the task of finding appropriate wave functionals to use for modeling the electron in the Stern-Gerlach experiment and calculating the time evolution of these wave functionals.  Although I am not aware of any analyses of this particular case within the field approach, there are some related accounts that one might build upon.  Taking a field approach for the electromagnetic field and a particle approach for the electron, \citet[sec.\ II.5]{bohm1987}; \citet[sec.\ 11.7]{bohmhiley} have shown how a wave functional initially representing a single incoming photon (modeled, for simplicity, using a scalar field) coupled with a wave function representing an electron in an atom will together evolve into a superposition of two distinct outcomes:\ (a) the photon being absorbed and ionizing the atom, and (b) the photon passing by without ionizing the atom.\footnote{See also \citet[sec.\ 4]{kaloyerou1994}.} \citeauthor{valentini1992} (\citeyear{valentini1992}, sec.\ 4.1, \citeyear{valentini1996}, pg.\ 54--55) discusses position measurements and the double-slit experiment for scalar field wave functionals.

\section{Conclusion}

In his famous ``Character of Physical Law'' lectures, \citet[ch.\ 6]{feynman} introduced ``the quantum mechanical view of nature'' by explaining how the double-slit experiment teaches us that electrons behave in some ways like bullets and in other ways like water waves.  In particular, they arrive at definite locations on the detector (like bullets) and display interference (like water waves).  Similarly, in the Stern-Gerlach experiment electrons behave in some ways like classical rigid bodies or point particles and in some ways like classical fields.  In particular, they arrive at definite locations on the detector (like classical rigid bodies or point particles) and they only hit the detector at two locations (like the classical Dirac field).  These behaviors are summarized in table \ref{table1}.\footnote{This table is modeled on Feynman's table distinguishing the behavior of bullets, water waves, and electrons \citep[figure 31]{feynman}.}

\setlength\tabcolsep{4 pt}
\begin{table}[h!]
\centering
\caption{A summary of which features of the Stern-Gerlach are explained in each of the different models of the electron that we have studied in sections \ref{rigidbodysection}-\ref{qftsection}.}
\begin{center}
\begin{small}
\begin{tabular}{ c|c|c|c|c|c|c }
& Classical  & Classical  & Non-Relativistic & Relativistic & Classical & Relativistic \\ 
  & Rigid & Point & Quantum & Quantum & Dirac & Quantum \\ 
  & Body & Particle & Particle & Particle & Field & Field  \\ \hline
Uniqueness & \ding{51} & \ding{51} & \ding{51} & \ding{51} & \ding{55} & \ding{51} \\ \hline
Discreteness & \ding{55} & \ding{55} & \ding{51} & \ding{51} & \ding{51} & \ding{51}  \\ 
\end{tabular}
\end{small}
\end{center}
\label{table1}
\end{table}

Looking at table \ref{table1}, it is clear that we need quantum physics to explain the results of the Stern-Gerlach experiment.  But, different classical models have different deficiencies.  If we focus on the rigid body and point particle theories in the first four columns, then the advantage of quantum physics is that it can explain the discreteness of outcomes (that electrons hit the detector in just two places).  If we focus instead on the field theories in the final two columns, then the advantage of quantum physics is different.  We need to move from classical to quantum field theory to explain the uniqueness of outcomes (that each electron hits the detector at a single location), and in that context we only get an explanation of uniqueness when we adopt a particular solution to the measurement problem.

\vspace*{12 pt}
\noindent
\textbf{Acknowledgments}
Thank you to Valia Allori, Jacob Barandes, Jeffrey Barrett, Mario Hubert, and the anonymous reviewers for helpful feedback and discussion.

\end{document}